\documentclass[11pt,a4paper]{article}
\usepackage{eurosym}
\usepackage{amssymb}
\usepackage{graphicx}
\usepackage{amsmath}
\usepackage{makeidx}
\usepackage{indentfirst}
\usepackage[T1]{fontenc}
\usepackage[utf8]{inputenc}
\usepackage{authblk}

\newcounter{resultnum}[section]
\newcounter{conclusionnum}[section]
\newcounter{conditionnum}[section]
\newcounter{conjecturenum}[section]
\newcounter{examplenum}[section]
\newcounter{exercisenum}[section]
\newcounter{lemmanum}[section]
\newcounter{notationnum}[section]
\newcounter{theoremnum}[section]
\newcounter{definitionnum}[section]
\newcounter{corollarynum}[section]
\newcounter{remarknum}[section]
\newcounter{propositionnum}[section]
\newcounter{acknowledgementnum}[section]
\newcounter{algorithmnum}[section]
\newcounter{axiomnum}[section]
\newcounter{casenum}[section]
\newcounter{claimnum}[section]
\newcounter{summarynum}[section]
\newcounter{problemnum}[section]
\begin{document}

\title{Broken Scale Invariance, Gravity Mass, and Dark Energy in\\
Modified Einstein Gravity with Two Measure Finsler Like Variables}
\date{March 30, 2021}
\author{ \vspace{.1 in} {\textbf{Panayiotis Stavrinos}}\thanks{%
email: pstavrin@math.uoa.gr } \\
{\small \textit{Department of Mathematics, National and Kapodistrian
University of Athens, \ Athens, 1584, Greece}} \vspace{.1 in} \\
\vspace{.1 in} \textbf{Sergiu I. Vacaru} \thanks{
emails: sergiu.vacaru@gmail.com ;\newline
\textit{Address for correspondence in 2021 as a visiting researcher at YF
CNU Ukraine:\ } Yu. Gagarin street, nr. 37/3, Chernivtsi, Ukraine, 58008} \\
{\small \textit{Physics Department, California State University at Fresno,
Fresno, CA 93740, USA; and }}\\
{\small \textit{Dep. Theoretical Physics and Computer Modelling, 101
Storozhynetska street, Chernivtsi, 58029, Ukraine}} }
\maketitle

\begin{abstract}
We study new classes of generic off--diagonal and diagonal cosmological
solutions for effective Einstein equations in modified gravity theories,
MGTs, with modified dispersion relations, MDRs, encoding possible violations
of (local) Lorentz invariance, LIVs. Such MGTs are constructed for actions
and Lagrange densities with two non--Riemannian volume forms (similar to two
measure theories, TMTs) and associated bimetric and/or biconnection
geometric structures. For conventional nonholonomic 2+2 splitting, we can
always describe such models in Finsler like variables, which is important
for elaborating geometric methods of constructing exact and parametric
solutions. Such Finsler two measure formulations of general relativity, GR,
and MGTs are considered for Lorentz manifolds and their (co) tangent bundles
and called, in brief, FTMT. Generic off-diagonal metrics solving
gravitational field equations in FTMTs are determined by generating
functions, effective sources and integration constants and characterized by
nonholonomic frame torsion effects. Restricting the class of integration
functions, we can extract torsionless and/or diagonal configurations and
model emergent cosmological theories with square scalar curvature, $R^2$,
when the global Weyl--scale symmetry is broken via nonlinear dynamical
interactions with nonholonomic constraints. In the physical Einstein-Finsler
frame, the constructions involve: (i) nonlinear re--parametrization
symmetries of the generating functions and effective sources; (ii) effective
potentials for the scalar field with possible two flat regions which allows
unified description of locally anisotropic and/or isotropic early universe
inflation related to acceleration cosmology and dark energy; (iii) there are
"emergent universes" described by off--diagonal and diagonal solutions for
certain nonholonomic phases and parametric cosmological evolution resulting
in various inflationary phases; (iv) we can reproduce in two measure
theories massive gravity effects. Finally, we study a reconstructing
procedure for reproducing off--diagonal FTMT and massive gravity
cosmological models as effective Einstein gravity or Einstein-Finsler
theories.

\vskip7pt

\end{abstract}

\newpage

\tableofcontents

\section{Introduction}

Modern cosmology has a very important task to provide a theoretical
description of many aspects of observable universe with exponential
expansion (inflation), particle creation, and radiation. We cite books \cite%
{cosmbooks1,cosmbooks2,cosmbooks3,cosmbooks4,cosmbooks5} on standard
cosmology \cite{stcosm1,stcosm2,stcosm3} and further developments. Then, on
acceleration cosmology \cite{accel1,accel2} and related dark energy and dark
matter physics, one can be considered a series of works on modified gravity
theories, MGTs, and cosmology \cite%
{mcosm1,mcosm2,mcosm3,bimmasv1,bimmasv2,vacaru18}. Another direction of
research is devoted to nonholonomic and Finsler like locally anisotropic
cosmological models \cite%
{fmgtcosm1,fmgtcosm2,fmgtcosm3,stavr1,stavr2,stavr3,stavr4}, see \cite%
{vacaru18a,bubuianu18a} for an axiomatic approach to
Finsler-Lagrange-Hamilton gravity theories. The physical community almost
accepted the idea that the Einstein gravity and standard particle physics
have to be modified in order to elaborate self--consistent quantum gravity
theories and describe existing experimental and observational data in modern
cosmology. In result, a number of MGTs and cosmological scenarios have been
elaborated during the last 20 years.

In a series of works \cite{vexactsol1,vexactsol1a,vexactsol2,vexactsol3,
bubuianu18,bubuianu19,bubuianu20, vacaru20}, see also references therein, it
was developed a geometric approach (the so--called anholonomic frame
deformation method, AFDM) for constructing exact and parametric solutions in
MGTs, general relativity, GR, and theory of nonholonomic geometric and
classical/ quantum information flows. Such solutions are with generic
off--diagonal metrics\footnote{%
such metrics can not be diagonalized via coordinate transforms in a finite
spacetime region; the solutions are, in general, with zero torsion
configurations; the Levi-Civita connection can be extracted by imposing
additional nonintegrable constraints (in physical and mathematical
literature, there are used also two equivalent terms, nonholonomic and/or
anholonomic conditions)} and generalized connections when their coefficients
depend on all spacetime coordinates via generating and integration
functions, for vacuum and non--vacuum configurations. One can be considered
effective and matter fields sources for possible Killing and non--Killing
symmetries and various types of commutative and noncommutative parameters
etc. For Finsler like modified gravity theories, FMGTs, the coefficients of
geometric and physical objects depend, in general, on (co) fiber velocity
(momentum) type coordinates. Following the AFDM, the geometric constructions
and variational calculus are preformed with respect to certain classes of
(adapted) nonholonomic frames for a formal splitting of spacetime dimension
in the form 2(3)+2+...+ and a well--defined geometrically "auxiliary" linear
connection which is convenient for performing, for instance, a deformation
quantization procedure, or for constructing exact and/or parametric
solutions. This allows to decouple the gravitational field equations in
MGTs, FMGTs, and GR, and geometric/information flow equations. Such
nonholonomic deformations of fundamental geometric objects determined by
distortions of nonlinear and linear connection structures were not
considered in other approaches with vierbeins (tetrads), 2+2 and/or 3+1
splitting, see standard textbooks on general relativity and exact solutions
\cite{texbgr1,texbgr2,texbgr3,texbgr4,texbgr5}. The methods elaborated by
other authors were successful only for generating exact solutions with two
and three Killing symmetries but do not provide a geometric/ analytic
formalism for a general decoupling of gravitational and matter field
equations. The surprising result is that such a decoupling is possible for
various classes of effective / modified Einstein equations and matter fields
which can be derived for certain physically motivated general assumptions in
MGTs.

Let us summarize most important ideas and methods developed in Refs. \cite%
{vacaru18,vacaru18a,bubuianu18a,vexactsol1,vexactsol1a,vexactsol2,vexactsol3, bubuianu18,bubuianu19, bubuianu20,vacaru20}%
:

\begin{description}
\item[a)] The (modified) Einstein equations with some effective and/or
matter field sources consist very sophisticate systems of nonlinear partial
derivative equations, PDEs. The bulk of most known and important physical
applications (of black hole, cosmological and other type solutions) were
elaborated for ansatz of metrics which can be diagonalized by certain
frame/coordinate transforms and when physically important systems of
nonlinear PDEs can be reduced to systems of decoupled nonlinear ODE
(ordinary differential equations). In such cases, the generated exact or
parametric solutions (i.e. integrals, with possible non-trivial topology,
singularities, of different smooth classes etc.) depend on one space, or
time, like coordinate, being determined by certain imposed symmetries (for
instance, spherical / axial ones, being invariant on some rotations, with
Lie algebras symmetries etc). The integration constants can be found in
explicit form considering certain symmetry / Cauchy / boundary / asymptotic
conditions. This way, there were constructed various classes of black/worm
hole and isotropic and anisotropic cosmological solutions.

\item[b)] The AFDM allows us to decouple and integrate physically important
systems of nonlinear PDEs in more general forms than in a) when the integral
varieties are parameterized not only by integration constants but also by
generating and integration functions subjected to nonholonomic constraints
and functional/ nonlinear dependence on sources and data for certain classes
of 'prime metrics and connections'. The resulting 'target' off--diagonal
metrics and generalized connections depend, in general, on all spacetime
coordinates. It is important to note that at the end we can impose
additional nonholonomic constraints and consider 'smooth' limits or various
type non-trivial topology and/or parametric transitions to Levi-Civita
configurations (with zero torsion) and/or diagonal metrics. This way we can
reproduce well known black hole / cosmological solutions etc., or which can
be with deformed horizons (for instance, with ellipsoid / toroid
symmetries), anisotropic polarized physical constants and, for instance,
imbedding into nontrivial gravitational vacuum configurations. Such new
classes of solutions can not be constructed if we impose from the very
beginning certain particular type ansatz for diagonalizable metrics, frames
of references and/or sources depending only on one spacetime coordinate.
This is an important property of nonlinear parametric physical systems
subjected to certain nonholonomic constraints. More general solutions with
geometric rich structure and various applications for a nonlinear
gravitational and matter fields dynamics can be found if we succeed to solve
directly certain generic nonlinear systems of PDEs not transformed into
systems of ODEs. Having constructed such general classes of solutions, one
might be analyzed limits to diagonal configurations and possible
perturbative effects. We "loose" the bulk of generic nonlinear solutions
with multi-variables if we consider from the very beginning certain
"simplified" ansatz for "higher--symmetries" resulting in ODEs.
\end{description}

Applying the AFDM as we explained above in paragraph b) and choosing
corresponding types of generating functions and integration functions and
constants, it is possible to model various MGTs and accelerating cosmology
effects by considering generic off--diagonal interactions and
re--parameterizations of generating functions and sources in effective
Einstein gravity. In the present paper, we shall elaborate on an unified
cosmological scenario for MGTs and GR with nonholonomic off--diagonal
interactions when effective Finsler like variables can be considered for a
2+2 splitting. In such an approach, both inflation and slowly accelerated
universe models are reproduced by exact solutions constructed following the
AFDM. In general, such solutions are inhomogeneous and with local
anisotropy. For a corresponding class of generating and integration
functions and for necessary type effective sources, we can model effective
scalar field potentials with anisotropy and limits to two flat regions. We
shall construct and study nonlinear parametric cosmological theories
generalizing the standard models based on
Friedman--Lama\^itre--Robertson--Walker, FLRW, diagonalizable configurations
derived for ODEs. The goal is to address the initial singularity problem and
to explain how two periods of exponential expansion with widely different
scales can be described via solutions of effective gravitational equations.

A well--known mechanism for generating accelerated expansion as a
consequence of vacuum energy can be performed in the context of scalar field
theory paradigm which is described by an effective potential $\ ^{ef}U$ with
flat regions. For such "slow roll" configurations of the vacuum field, the
kinetic energy terms are small and the resulting energy--momentum tensor is
of type $T_{\mu \nu }\simeq g_{\mu \nu }\ ^{ef}U$. If the potential $\
^{ef}U $ contains contributions of some modified gravity terms (two
measures, massive gravity and other ones), we can analyse possible effects
of such terms in the inflationary phase. But this is not enough for
elaborating a theory of modern cosmology with acceleration and dark energy
and dark matter contributions. Theoreticians developed different
quintessential, $k$-essence and "variable gravity" inflation scenarios \cite%
{quint1,quint2,quint3,quint4,quint5,quint6,quint7,quint8} and $f(R)$
modified models, in general, with contributions from massive gravity,
Finsler like theories, bi--metrics and bi--connections and/or generic
off--diagonal interactions, see \cite{bimmasv3,bimmasv4,bimmasv1,bimmasv2,
fmgtcosm1,fmgtcosm2,vacaru18a,bubuianu18a,bubuianu18} and references therein.

The solutions with anisotropies and flat regions can be used for
speculations on the phase that proceeds the inflation and may explain both a
non--singular origin of universe and the early universe evolution. This is
similar to the concept of "emergent universe" which was considered with the
aim to solve the problem of initial singularity including the singularity
theorems for inflationary cosmology driven by scalar field \cite%
{emergent1,emergent2,emergent3,emergent4,emergent5,emergent6}. In our
approach with solutions constructed following the AFDM, the universe does
not start as a static Einstein universe but as a parametric effective one
when the scalar field rolls with an almost constant speed for a
non--singular configuration with small anisotropies.

Let us briefly motivate and explain the origins of the present work. The
main ideas and methods on constructing generic off--diagonal solutions in
MGTs comes from Refs. \cite{vexactsol1,vexactsol1a,vexactsol2,vexactsol3,
bubuianu18,bubuianu19}. In articles \cite{mcosm1,mcosm2,mcosm3,
bimmasv1,bimmasv2,vacaru18,fmgtcosm1,fmgtcosm2,fmgtcosm3,bimmasv1,bimmasv2},
there were considered various examples when the gravitational and matter
field equations in MGTs can be re--defined and solved as certain effective/
generalized Einstein equations or their Finsler like modifications. A series
of papers \cite{guend1a,guend1b,guend1c,guend1d,guend1e} is devoted to a new
class of modified--measure gravity--matter theories containing different
terms in the pertinent Lagrangian action, for instance, one with a
non-Riemannian integration measure and another with standard Riemannian
integration measure. In brief, we shall call such models as two measure
theories, TMTs, of gravity. In a more general case, there are considered two
non--metric densities \cite{guend2}. An important feature of such theories
is that the constructions are with global Weyl--scale invariance and further
dynamical breaking. In particular, the second action term is the standard
Riemannian integration measure containing a Weyl--scale symmetry preserving $%
R^{2},$ or more general $f(R)$ terms, which in this work may encode
modifications from massive gravity, bi--metric and bi--connection theories.
The latter formalism and geometrization of such TMTs allow to represent the
corresponding gravitational field equations as certain effective Einstein
equations in nonholonomic variables, see various applications in modern
cosmology, (super) string/ brane theories, non--Abelian confinement etc.
\cite{guend3a,guend3b}. The main goal of this article is to develop the AFDM
for generating exact solutions in TMTs formulated in nonholonomic and
Einstein-Finsler variables, see also a partner work \cite{rajpootvacaru},
and analyze possible implications in modern cosmology and f dark energy and
dark matter physics.

The work is organized as follows. Geometric preliminaries on nonholonomic
Lorentz manifolds and relativistic Lagrange-Finsler spaces are provided in
section \ref{perlimsect}. Then, in section \ref{mgttmt}, we formulate a
geometric approach to MGT cosmology in the framework of TMT with
nonholonomic variables and effective Einstein-Finsler gravity theories. We
apply the AFDM for constructing generic off--diagonal cosmological solutions
in various MGTs in section \ref{scosmsol}. Cosmological models with locally
anisotropic effective scalar potentials and two flat regions are studied in
section \ref{stwoflat}. We devote the section \ref{smassgrfromtmt} to
formulation of certain conditions when modified massive gravity can be
reproduced as TMTs and effective GR theories, with nonholonomic Finsler like
variables, and speculate on reconstructing procedure for such massive
gravity cosmological models. Finally, we provide a discussion and
conclusions in section \ref{sconcl}.

\section{Nonholonomic variables and (modified) Einstein and Lagrange-Finsler
equations}

\label{perlimsect}In this section, we outline some necessary results from
the geometry of four dimensional, 4-d, Lorentz manifolds with so-called
canonical nonholonomic variables which can be transforms in Finsler-Lagrange
like variables. The motivation to consider canonical variables is that they
allow to prove certain general decoupling and integration properties of
gravitational field equations in MGTs and GR, but Finsler-Lagrange like
variables and associated almost symplectic structures which can be used for
deformation and other type quantization procedures of gravity theories.
Proofs and details can be found, for instance, in \cite%
{vacaru18a,bubuianu18a}.

\subsection{Geometric objects and GR and MGTs in nonholonomic variables}

Let us consider a 4-d pseudo--Riemannian manifold $V$ defined by a metric
structure%
\begin{equation}
g=g_{\alpha \beta }(u^{\gamma })du^{\alpha }\otimes du^{\beta }
\label{offdm}
\end{equation}%
of signature $(+,+,+-),$ with local coordinates $u=\{u^{\gamma }\},$ where
indices $\alpha ,\beta ,\gamma ,...$ run values $1,2,3$ and (for the time
like coordinate) $4$. The Einstein summation rule on up/low repeating
indices is applied if it will be not stated the contrary. For a
corresponding causality structure combining locally the postulates of the
special relativity theory, the principle of equivalence etc. (see a review
of axiomatic appraoches in GR and Finsler like modified theories in \cite%
{fmgtcosm1,vacaru18a,bubuianu18a}) such a curved spacetime is called a
Lorentz manifold. In this work, we study generalizations of geometric and
gravitational and cosmological models when certain nonholonomic
(nonintegrable, anholonomic) distributions and related bimeasure structures,
and Lagrangians for MGTs, are considered on $V.$

On a curved spacetime $V,$ we can always introduce a nonholonomic $2+2$
splitting which is determined by a non--integrable distribution
\begin{equation*}
\mathbf{N}:T\mathbf{V}=h\mathbf{V}\oplus v\mathbf{V,}
\end{equation*}%
where $T\mathbf{V}$ is the tangent bundle of $V,$ the Withney sum $\oplus $
defines a conventional splitting into horizontal (h), $h\mathbf{V,}$ and
vertical (v), $v\mathbf{V,}$ subspaces. In local cooridnates
\begin{equation}
\mathbf{N=}N_{i}^{a}(x^{k},y^{b})dx^{i}\otimes \partial /\partial y^{a},
\label{ncon}
\end{equation}%
states a nonlinear connection, N--connection structure. For such a
N--connection decomposition, the indices and coordinates split in the form $%
u=(x,y),$ or $u^{\alpha }=(x^{i},y^{a}),$ for $x=\{x^{i}\}$ and $%
y=\{y^{a}\}, $ with $i,j,k,...=1,2$ and $a,b,c,...=3,4,$ which is
respectively adapted to a nonholonomic $2+2$ splitting. The data $(\mathbf{%
V,N})$ define a nonholonomic manifold with a prescribed fibered structure
described locally by fiber like coordinates $y^{a}$.

In our works, there are used "boldface" symbols in order to emphasize that
certain geometric/ physical objects are defined for spaces enabled with a
2+2 splitting determined by a N--connection structure. On pseudo--Riemannian
manifolds, to introduce a N--connection with a 2+2 splitting is equivalent
to the convention that there are used certain subclasses of local
(N--adapted) bases $\mathbf{e}_{\mu }=(\mathbf{e}_{i},e_{a})$ and their
duals $\mathbf{e}^{\nu }=(e^{j},\mathbf{e}^{b}),$ where
\begin{equation}
\mathbf{e}_{i}=\frac{\partial }{\partial x^{i}}-N_{i}^{c}\frac{\partial }{%
\partial y^{c}},e_{a}=\partial _{a}=\frac{\partial }{\partial y^{a}}%
\mbox{
and }e^{j}=dx^{j},\mathbf{e}^{b}=dy^{b}+N_{k}^{b}dx^{k}.  \label{nadaptb}
\end{equation}%
Such frames are called nonholonomic because they satisfy, in general, the
relations%
\begin{equation*}
\lbrack \mathbf{e}_{\alpha },\mathbf{e}_{\beta }]=\mathbf{e}_{\alpha }%
\mathbf{e}_{\beta }-\mathbf{e}_{\beta }\mathbf{e}_{\alpha }=W_{\alpha \beta
}^{\gamma }\mathbf{e}_{\beta },
\end{equation*}%
with nontrivial anholonomy coefficients $W_{ia}^{b}=\partial
_{a}N_{i}^{b},W_{ji}^{b}=\Omega _{ij}^{b}=\mathbf{e}_{j}(N_{i}^{b})-\mathbf{e%
}_{i}(N_{j}^{b})$. For zero $W$--coefficients, we get holonomic bases which
allows to consider coordinate transforms $\mathbf{e}_{\alpha }\rightarrow
\mathbf{\partial }_{\alpha }$ and $\mathbf{e}^{\beta }\rightarrow du^{\beta
}.$

On any manifold $V$ and its tangent and cotangent bundle, there are also
possible general vierbein (tetradic) transforms $e_{\alpha }=e_{\ \alpha }^{%
\underline{\alpha }}(u)\partial /\partial u^{\underline{\alpha }}$ and $%
e^{\beta }=e_{\ \underline{\beta }}^{\beta }(u)du^{\underline{\beta }}$,
where the coordinate indices are underlined in order to distinguish them
from arbitrary abstract ones and the matrix $e_{\ \underline{\beta }}^{\beta
}$ is inverse to $e_{\ \alpha }^{\underline{\alpha }}$ for orthonormalized
bases. We do not use boldface symbols for such transforms because an
arbitrary decomposition (we can consider as particular cases certain diadic
2+2 splitting) is not adapted to a N--connection structure.

With respect to N--adapted bases, we shall say that a vector, a tensor and
other geometric objects are represented correspondingly as a distinguished
vector (d--vector), a distinguished tensor (d--tensor) and distinguished
objects (d--object). Using frame transforms $\mathbf{g}_{\alpha \beta }=e_{\
\alpha }^{\alpha ^{\prime }}e_{\ \beta }^{\beta ^{\prime }}g_{\alpha
^{\prime }\beta ^{\prime }},$ any metric $g$ (\ref{offdm}) on $\mathbf{V}$
can be written in N--adapted from as a distinguished metric (in brief,
d--metric)
\begin{equation}
\mathbf{g}=\mathbf{g}_{\alpha \beta }\mathbf{e}^{\alpha }\otimes \mathbf{e}%
^{\beta }=g_{ij}(u)dx^{i}\otimes dx^{j}+g_{ab}(u)\mathbf{e}^{a}\otimes
\mathbf{e}^{b}.  \label{dm}
\end{equation}%
In brief, such a h-v decomposition of a metric structure is parameterized in
the form \newline
$\mathbf{g}=\left(h\mathbf{g}=\{g_{ij}\},v\mathbf{g}=\{g_{ab}\}\right).$

On nonholonomic manifolds, we can work with a subclass of linear connections
$\mathbf{D}=(h\mathbf{D},v\mathbf{D}),$ called distinguished connections,
d--connections, preserving under parallelism the N--connection splitting. A
d-connection is determined by its coefficients $\mathbf{\Gamma }_{\ \beta
\gamma }^{\alpha }=\{L_{\ jk}^{i},\acute{L}_{\ bk}^{a},\acute{C}_{\
jc}^{i},C_{\ bc}^{a}\}$ computed with respect to a N-adapted base (\ref%
{nadaptb}). Not adapted to N-connections linear connections structures can
be also considered but they do not preserve under parallelism (\ref{ncon})
and satisfy other types transformation laws under frame/coordinate
transforms.

For any d--vectors $\mathbf{X}$ and $\mathbf{Y,}$ we can define in standard
form, correspondingly, the torsion d--tensor, $\mathcal{T},$ the
nonmetricity d--tensor, $\mathcal{Q},$ and the curvature d--tensor, $%
\mathcal{R},$ of a $\mathbf{D}$, which (in general) do not depend on $%
\mathbf{g}$ and/or $\mathbf{N}$. The formulas are
\begin{equation}
\mathcal{T}(\mathbf{X,Y}):=\mathbf{D}_{\mathbf{X}}\mathbf{Y}-\mathbf{D}_{%
\mathbf{Y}}\mathbf{X}-[\mathbf{X,Y}],\mathcal{Q}(\mathbf{X}):=\mathbf{D}_{%
\mathbf{X}}\mathbf{g},\mathcal{R}(\mathbf{X,Y}):=\mathbf{D}_{\mathbf{X}}%
\mathbf{D}_{\mathbf{Y}}-\mathbf{D}_{\mathbf{Y}}\mathbf{D}_{\mathbf{X}}-%
\mathbf{D}_{\mathbf{[X,Y]}}.  \label{dgeomob}
\end{equation}%
In N--adapted coefficients labeled by h- and v-indices, such geometric
d-objects are parameterized respectively
\begin{eqnarray*}
\mathcal{T}&=&\{\mathbf{T}_{\ \alpha \beta }^{\gamma }=\left( T_{\
jk}^{i},T_{\ ja}^{i},T_{\ ji}^{a},T_{\ bi}^{a},T_{\ bc}^{a}\right) \},%
\mathcal{Q}=\mathbf{\{Q}_{\ \alpha \beta }^{\gamma }\}, \\
\mathcal{R}&=&\mathbf{\{R}_{\ \beta \gamma \delta }^{\alpha }=\left( R_{\
hjk}^{i},R_{\ bjk}^{a},R_{\ hja}^{i},R_{\ bja}^{c},R_{\ hba}^{i},R_{\
bea}^{c}\right) \}.
\end{eqnarray*}%
Such coefficients can be computed in explicit form by introducing $\mathbf{X}%
=\mathbf{e}_{\alpha }$ and $\mathbf{Y}=\mathbf{e}_{\beta },$ see (\ref%
{nadaptb}), and coefficients of a d-connection$\mathbf{D}=\{\mathbf{\Gamma }%
_{\ \alpha \beta }^{\gamma }\}$ into formulas (\ref{dgeomob}).

To elaborate the AFDM is convenient to work with two "preferred" linear
connections:\ the Levi--Civita connection $\nabla $ and the canonical
d--connection $\widehat{\mathbf{D}}$. Both such connections are completely
defined by a metric structure $\mathbf{g}$ following the conditions
\begin{equation}
\mathbf{g}\rightarrow \left\{
\begin{array}{cc}
\mathbf{\nabla :} & \ ^{\nabla }\mathcal{Q}=0\mbox{ and }\ ^{\nabla }%
\mathcal{T}=0; \\
\widehat{\mathbf{D}}: & \widehat{\mathcal{Q}}=0\mbox{ and }\ h\widehat{%
\mathcal{T}}=0,v\widehat{\mathcal{T}}=0,%
\end{array}%
\right.  \label{lcon}
\end{equation}%
where the left label $\nabla $ is used for the geometric objects determined
by the Levi--Civita, LC, connection. It should be noted here that the
N--adapted coefficients of the torsion $\widehat{\mathcal{T}}$ are not zero
for the case of mixed $h$- and $v$-coefficients computed with respect to
N--adapted frames (conventionally, we can write this as $hv\widehat{\mathcal{%
T}}\neq 0,$ with some nontrivial N--adapted coefficients from the subset $%
\{T_{\ ja}^{i},T_{\ ji}^{a},T_{\ bi}^{a}\}).$ Such a torsion $\widehat{%
\mathcal{T}}$ is completely determined by the coefficients of $\mathbf{N}$
and $\mathbf{g}$ (in coordinate frames, such values determine certain
generic off--diagonal terms $g_{\alpha \beta }$ which can not be
diagonalized in a finite spacetime region $U\subset \mathbf{V}$ by
coordinate transforms). We can consider a distortion relation%
\begin{equation*}
\widehat{\mathbf{D}}=\nabla +\widehat{\mathbf{Z}},
\end{equation*}%
when both linear connections and the distortion tensors $\widehat{\mathbf{Z}}
$ are completely defined by the geometric data $(\mathbf{g},\nabla ),$ or
(in nonholonomic variables) by $(\mathbf{g},\mathbf{N},\widehat{\mathbf{D}}%
). $

Contracting the indices of a canonical Riemann d--tensor of $\widehat{%
\mathbf{D}},$ $\widehat{\mathcal{R}}\mathbf{=}\mathbf{\{}\widehat{\mathbf{R}}%
_{\ \beta \gamma \delta }^{\alpha }\},$ we construct a respective canonical
Ricci d--tensor, $\widehat{\mathcal{R}}ic=\{\widehat{\mathbf{R}}_{\alpha
\beta }:=\widehat{\mathbf{R}}_{\ \alpha \beta \gamma }^{\gamma }\}.$ The
corresponding nontrivial N--adapted coefficients are
\begin{equation}
\widehat{\mathbf{R}}_{\alpha \beta }=\{\widehat{R}_{ij}:=\widehat{R}_{\
ijk}^{k},\ \widehat{R}_{ia}:=-\widehat{R}_{\ ika}^{k},\ \widehat{R}_{ai}:=%
\widehat{R}_{\ aib}^{b},\ \widehat{R}_{ab}:=\widehat{R}_{\ abc}^{c}\},
\label{dricc}
\end{equation}%
when the scalar curvature is computed $\ $%
\begin{equation*}
\widehat{R}:=\mathbf{g}^{\alpha \beta }\widehat{\mathbf{R}}_{\alpha \beta
}=g^{ij}\widehat{R}_{ij}+g^{ab}\widehat{R}_{ab}.
\end{equation*}%
It should be noted that, in general, $\widehat{\mathbf{R}}_{\alpha \beta
}\neq \widehat{\mathbf{R}}_{\beta \alpha },$ even such a tensor is symmetric
for the LC-connection, $R_{\alpha \beta }=R_{\beta \alpha }.$ This a
nonholonomic deformation and nonholonomic frame effect.

We can introduce the Einstein d--tensor%
\begin{equation*}
\widehat{\mathbf{E}}_{\alpha \beta }:=\widehat{\mathbf{R}}_{\alpha \beta }-%
\frac{1}{2}\mathbf{g}_{\alpha \beta }\ \widehat{R}
\end{equation*}%
and consider an effective Lagrangian $\ \widehat{L}$ for which the
stress--energy momentum tensor, $\widehat{\mathbf{T}}_{\alpha \beta }$, is
defined by an N--adapted (with respect to $\mathbf{e}_{\beta }$ and $\mathbf{%
e}^{\alpha })$ variational calculus on a nonholonomic manifold $(\mathbf{g},%
\mathbf{N},\widehat{\mathbf{D}})$,
\begin{equation}
\widehat{\mathbf{T}}_{\alpha \beta }=-\frac{2}{\sqrt{|\mathbf{g}_{\mu \nu }|}%
}\frac{\delta (\sqrt{|\mathbf{g}_{\mu \nu }|}\ \widehat{\mathcal{L}})}{%
\delta \mathbf{g}^{\alpha \beta }}.  \label{estrt}
\end{equation}%
Following geometric principles, we can postulate the Einstein equations in
GR for the data $(\mathbf{g,}\widehat{\mathbf{D}}),$ and/or to re--write
them equivalently for the data$(\mathbf{g},\nabla )$ if additional
nonholonomic constraints for zero torsion are imposed,
\begin{eqnarray}
\widehat{\mathbf{R}}_{\ \beta \gamma } &=&\widehat{\Upsilon }_{\ \beta
\gamma },  \label{cdeinst} \\
\mbox{ and \quad }\widehat{\mathcal{T}} &=&0,\mbox{ additional condition for}%
\nabla .  \label{lccond}
\end{eqnarray}%
In general, the condition $\widehat{\mathbf{D}}_{\mid \widehat{\mathcal{T}}%
=0}=\nabla $ may not have a smooth limit and such an equation can be
considered as a nonholonomic or parametric constraint. Here we note that the
source
\begin{equation*}
\widehat{\Upsilon }_{\ \beta \gamma }:=\varkappa (\widehat{\mathbf{T}}%
_{\alpha \beta }-\frac{1}{2}\mathbf{g}_{\alpha \beta }\widehat{T})
\end{equation*}%
is computed with the trace $\widehat{T}:=\mathbf{g}^{\alpha \beta }\widehat{%
\mathbf{T}}_{\alpha \beta }$ and $\varkappa $ should be determined by the
Newton constant $^{New}G$ as in GR if we wont to study limits to the
Einstein gravity theory. In this work, we shall use the units when $%
^{New}G=1/16\pi $ and the Planck mass $^{Pl}M=(8\pi ^{New}G)^{-1/2}=\sqrt{2}$%
. If we do not impose the LC-conditions (\ref{lccond}), the system of
nonholonomic nonlinear PDEs (\ref{cdeinst}), and similar higher dimension
ones, for instance, with noncommutative and/or supersymmetric variables can
be considered in various classes of MGTs, Finsler-Lagrange gravity etc.

The values $\widehat{\mathcal{R}},\widehat{\mathcal{R}}ic$ and $\ \widehat{R}
$\ for the canonical d--connection $\widehat{\mathbf{D}}$ are different from
the similar ones, $\mathcal{R},\mathcal{R}ic$ and $R,$ computed for the
LC--connection $\nabla $. Nevertheless, both classes of such fundamental
geometric objects are related via distorting relations derived in a unique
form for a given metric structure and N--connection splitting. There are at
least two priorities to work with $\widehat{\mathbf{D}}$ instead of $\nabla
. $ The first one is that we can find solutions for generalized gravity
theories with nontrivial torsion. The second priority is that the equations (%
\ref{cdeinst}) decouple in very general forms with respect to certain
classes of N--adapted frames. The basic idea of the AFDM is to write the
Lagrange densities and the resulting fundamental gravitational and matter
field equations in terms of such nonholonomic variables which allows us to
decouple and solve nonlinear systems of PDEs. This can not be done if we use
from the very beginning the LC--connection $\nabla $. It is not a
d--connection, does not preserve under general transforms the $h$- and $v$%
--splitting and the condition of zero torsion, $\ ^{\nabla }\mathcal{T}=0$
and does not allow to decouple the equations in general forms. Working with $%
\widehat{\mathbf{D}}$, we introduce certain "flexibility" in order apply a
corresponding geometric techniques for integration PDEs. In such cases, we
do not make additional assumptions on particular cases for ansatz and
connections transforming the fundamental field equations into nonlinear
systems of ODEs. Having defined a quite general class of solutions expressed
via generating functions and integration functions and constants, we can
impose additional nonholonomic constraints (\ref{lccond}) which allows to
extract LC--configurations. This way, we can construct in explicit form new
classes of exact solutions in GR and MGTs both in $(\mathbf{g},\nabla )$ and
$(\mathbf{g},\mathbf{N},\widehat{\mathbf{D}})$ variables.

\subsection{Finsler-Lagrange variables in GR and MGTs}

On a 4-d/ Lorentz manifold $V,$ we can introduce always Finsler like
variables considering a conventional 2+2 splitting of coordinates $%
u^{\alpha}=(x^{i},y^{a})$ for a nonholonmic fiber structure where $%
y=\{y^{a}\},$ for $a=3,4,$ are treated as effective fiber coordinates (which
are analogous to velocity ones in theories on tangent bundles). This way we
elaborate a toy model of relativistic Finsler-Lagrange geometry. Let us
explain how such constructions provide examples of above formulated
nonholonomic models of (pseudo) Riemannian geometry. A fundamental function
(equivalently, generating function) $V\ni (x,y)\rightarrow L(x,y)\in \mathbb{%
R},$ i.e. a real valued function (in brief, called an effective Lagrangian,
or a Lagrange density) which is differentiable on $\widetilde{V}:=V/\{0\},$
for $\{0\}$ being the null section of $V,$ and continuous on the null
section of $\pi :V\rightarrow hV.$ A relativistic 4-d model of a fibered
effective Lagrange space $L^{3,1}=(V,L(x,y))$ is determined by a prescribed
regular Hessian metric (equivalently, v-metric)
\begin{equation}
\widetilde{g}_{ab}(x,y):=\frac{1}{2}\frac{\partial ^{2}L}{\partial
y^{a}\partial y^{b}}  \label{hessls}
\end{equation}%
is non-degenerate, i.e. $\det |\widetilde{g}_{ab}|\neq 0,$ and of constant
signature. Non-regular configurations can be studied as special cases.

The non-Riemannian total phase space geometries are characterized by
nonlinear quadratic line elements
\begin{equation}
ds_{L}^{2}=L(x,y).  \label{nqed}
\end{equation}%
We can elaborate on geometric and physical theories with a spacetime enabled
with a nonholonomic frame and metric and (non) linear connection structures
determined by a nonlinear quadratic line element (\ref{nqed}) and related
v-metric (\ref{hessls}). The geometric objects on $L^{3,1}$ will be labeled
by a tilde "\symbol{126}" (for instance, $\widetilde{g}_{ab}$) if they are
defined canonically by an effective Lagrange generating function. We write $%
\widetilde{L}^{3,1}$ with tilde in order to emphasize that $V$ is enabled
with an effective relativistic Lagrange structure and respective
nondegenerate Hessian.

The dynamics of a probing point particle in $\widetilde{L}^{3,1}$ is
described by Euler-Lagrange equations, $\ $%
\begin{equation*}
\frac{d}{d\tau }\frac{\partial \widetilde{L}}{\partial y^{i}}-\frac{\partial
\widetilde{L}}{\partial x^{i}}=0.
\end{equation*}%
These equations are equivalent to the \textit{nonlinear geodesic
(semi-spray) equations}
\begin{eqnarray}
\frac{d^{2}x^{i}}{d\tau ^{2}}+2\widetilde{G}^{i}(x,y) &=&0,  \label{ngeqf} \\
\mbox{ for } \widetilde{G}^{i} &=& \frac{1}{2}\widetilde{g}^{ij}(\frac{%
\partial ^{2} \widetilde{L}}{\partial y^{i}}y^{k}-\frac{\partial \widetilde{L%
}}{\partial x^{i}}),\,  \notag
\end{eqnarray}%
where $\widetilde{g}^{ij}$ is inverse to $\ \widetilde{g}_{ab}$ (\ref{hessls}%
). This way we define a canonical Lagrange N-connection structure

\begin{equation}
\ \ \ \widetilde{\mathbf{N}}=\left\{ \widetilde{N}_{i}^{a}:=\frac{\partial
\widetilde{G}}{\partial y^{i}}\right\} ,  \label{lagnc}
\end{equation}%
determining an effective Lagrange N-splitting $\widetilde{\mathbf{N}}:\
TV=hV\oplus vV,$ similar to (\ref{ncon}). Using $\widetilde{N}_{i}^{a} $
from (\ref{lagnc}), we define effective Lagrange N--adapted (co) frames
\begin{equation}
\widetilde{\mathbf{e}}_{\alpha }=(\widetilde{\mathbf{e}}_{i}=\frac{\partial
}{\partial x^{i}}-\widetilde{N}_{i}^{a}(x,y)\frac{\partial }{\partial y^{a}}%
,e_{b}=\frac{\partial }{\partial y^{b}}) \mbox{ and } \widetilde{\mathbf{e}}%
^{\alpha }=(\widetilde{e}^{i}=dx^{i},\widetilde{\mathbf{e}}^{a}=dy^{a}+%
\widetilde{N}_{i}^{a}(x,y)dx^{i}).  \label{cnddapb}
\end{equation}%
Such $\widetilde{N}$-adapted frames can be considered as results of certain
vierbein (frame, for 4-d, tetradic) transforms of type $e_{\alpha }=e_{\
\alpha }^{\underline{\alpha }}(u)\partial /\partial u^{\underline{\alpha }}$
and $e^{\beta }=e_{\ \underline{\beta }}^{\beta }(u)du^{\underline{\beta }}.$%
\footnote{%
We can underline the local coordinate indices in order to distinguish them
from arbitrary abstract ones; the matrix $e_{\ \underline{\beta }}^{\beta }$
is inverse to $e_{\ \alpha }^{\underline{\alpha }}$ for orthonormalized
bases.}

We can also consider frame transforms $e_{\alpha }=e_{\ \alpha }^{\alpha
^{\prime }}(u)e_{\alpha ^{\prime }},$ when $\widetilde{g}_{ij}=e_{\
i}^{i^{\prime }}e_{\ k}^{j^{\prime }}\widetilde{g}_{i^{\prime }j^{\prime }}$
and $\widetilde{g}_{ab}=e_{\ a}^{a^{\prime }}e_{\ b}^{b^{\prime }}\widetilde{%
g}_{a^{\prime }b^{\prime }}$ for $\widetilde{g}_{i^{\prime }j^{\prime }}$
and $\widetilde{g}_{a^{\prime }b^{\prime }}$ being of type (\ref{hessls}),
define the respective h- and v-components of a d-metric of signature $%
(+++-). $ In result, we can construct a relativistic Sasaki type d-metric
structure%
\begin{equation}
\widetilde{\mathbf{g}}=\widetilde{\mathbf{g}}_{\alpha \beta }(x,y)\widetilde{%
\mathbf{e}}^{\alpha }\mathbf{\otimes }\widetilde{\mathbf{e}}^{\beta }=%
\widetilde{g}_{ij}(x,y)e^{i}\otimes e^{j}+\widetilde{g}_{ab}(x,y)\widetilde{%
\mathbf{e}}^{a}\otimes \widetilde{\mathbf{e}}^{a}.\   \label{dmds}
\end{equation}%
Using respective frame transforms $\mathbf{g}_{\alpha ^{\prime }\beta
^{\prime }}=e_{\ \alpha ^{\prime }}^{\alpha }e_{\ \beta ^{\prime }}^{\beta }%
\widetilde{\mathbf{g}}_{\alpha \beta }$ and $\mathbf{g}_{\alpha ^{\prime
}\beta ^{\prime }}=e_{\ \alpha ^{\prime }}^{\underline{\alpha }}e_{\ \beta
^{\prime }}^{\underline{\beta }}g_{\underline{\alpha }\underline{\beta }}$,
such an effective Lagrange-Sasaki can be represented as a general d-metric (%
\ref{dm}), or equivalently, as a off-diagonal metric (\ref{offdm}),
\begin{equation*}
\mathbf{g}=\mathbf{g}_{\alpha ^{\prime }\beta ^{\prime }}(x,y)\mathbf{e}%
^{\alpha ^{\prime }}\mathbf{\otimes e}^{\beta ^{\prime }}=g_{\underline{%
\alpha }\underline{\beta }}(x,y)du^{\underline{\alpha }}\mathbf{\otimes }du^{%
\underline{\beta }},
\end{equation*}%
where
\begin{equation}
g_{\underline{\alpha }\underline{\beta }}=\left[
\begin{array}{cc}
g_{ij}(x)+g_{ab}(x,y)N_{i}^{a}(x,y)N_{j}^{b}(x,y) & g_{ae}(x,y)N_{j}^{e}(x,y)
\\
g_{be}(x,y)N_{i}^{e}(x,y) & g_{ab}(x,y)%
\end{array}%
\right] .  \label{offn}
\end{equation}%
Parameterizations of type (\ref{offn}) for metrics are considered in
Kaluza--Klein theories but in our approach the N-coefficients are determined
by a general or Lagrange N-connection structure.

The Lagrange N--connections $\widetilde{\mathbf{N}}$ defines an almost
complex structure $\widetilde{\mathbf{J}}.$ Such a linear operator $%
\widetilde{\mathbf{J}}$ acts on$\ \mathbf{e}_{\alpha }=(\ \mathbf{e}%
_{i},e_{b})$ using formulas $\ ^{\shortmid }\widetilde{\mathbf{J}}(\mathbf{e}%
_{i})=-e_{n+i}$ and $\widetilde{\mathbf{J}}(e_{n+i})=\mathbf{e}_{i},$ and
defines globally an almost complex structure $\widetilde{\mathbf{J}}\mathbf{%
\circ }\widetilde{\mathbf{J}}=$ $-\mathbf{\ I,}$ where $\mathbf{I}$ is the
unity matrix. We note that $\widetilde{\mathbf{J}}$ is a (pseudo) almost
complex structure only for a (pseudo) Euclidean signature. There are omitted
tildes and written, for instance, $\mathbf{J}$ for arbitrary frame/
coordinate transforms.

A Lagrange Neijenhuis tensor field is determined by a Lagrange generating
function introduced as the curvatures of a respective N--connection,
\begin{equation}
\widetilde{\mathbf{\Omega }}\mathbf{(}\widetilde{\mathbf{X}}\mathbf{,}%
\widetilde{\mathbf{Y}}):=\mathbf{-[\widetilde{\mathbf{X}}\mathbf{,}%
\widetilde{\mathbf{Y}}]+[\widetilde{\mathbf{J}}\widetilde{\mathbf{X}},%
\widetilde{\mathbf{J}}\widetilde{\mathbf{Y}}]-\widetilde{\mathbf{J}}[%
\widetilde{\mathbf{J}}\widetilde{\mathbf{X}},\widetilde{\mathbf{Y}}]-%
\widetilde{\mathbf{J}}[\widetilde{\mathbf{X}},\widetilde{\mathbf{J}}%
\widetilde{\mathbf{Y}}],}  \label{neijta}
\end{equation}%
for any d--vectors $\mathbf{X,}$ $\mathbf{Y.}$ Such formulas can be written
without tilde values if there are considered arbitrary frame transforms. In
local form, a N--connection is characterized by such coefficients of (\ref%
{neijta}) (i.e. the coefficients of a N--connection curvature):
\begin{equation}
\Omega _{ij}^{a}=\frac{\partial N_{i}^{a}}{\partial x^{j}}-\frac{\partial
N_{j}^{a}}{\partial x^{i}}+N_{i}^{b}\frac{\partial N_{j}^{a}}{\partial y^{b}}%
-N_{j}^{b}\frac{\partial N_{i}^{a}}{\partial y^{b}}.  \label{neijtc}
\end{equation}%
An almost complex structure $\mathbf{J}$ transforms into a standard complex
structure for the Euclidean signature if $\mathbf{\Omega }=0.$

Using the Lagrange d-metric $\widetilde{\mathbf{g}}$ and d-operator $\mathbf{%
\widetilde{\mathbf{J}},}$ we can define the almost symplectic structure $%
\widetilde{\theta }:=\widetilde{\mathbf{g}}(\widetilde{\mathbf{J}}\cdot
,\cdot ).$ Then, we can construct canonical d--tensor fields defined by $%
L(x,y)$ and N-adapted respectively to $\widetilde{N}_{i}^{a}$ (\ref{lagnc})
and $\widetilde{\mathbf{e}}_{\alpha }=(\widetilde{\mathbf{e}}_{i},e_{b})$ (%
\ref{cnddapb}):%
\begin{eqnarray}
\widetilde{\mathbf{J}} &=&-\delta _{i}^{a}e_{a}\otimes e^{i}+\delta _{a}^{i}%
\widetilde{\mathbf{e}}_{i}\otimes \widetilde{\mathbf{e}}^{a}\
\mbox{ the
almost complex structure };  \label{aux21} \\
\ \widetilde{\mathbf{P}} &=&\widetilde{\mathbf{e}}_{i}\otimes
e^{i}-e_{a}\otimes \widetilde{\mathbf{e}}^{a}%
\mbox{ almost product structure
};  \notag \\
\widetilde{\theta } &=&\widetilde{g}_{aj}(x,y)\widetilde{\mathbf{e}}%
^{a}\wedge e^{i}\mbox{ almost symplectic structure }.  \notag
\end{eqnarray}%
We can define the Cartan-Lagrange d-connection $\widetilde{\mathbf{D}}=(h%
\widetilde{\mathbf{D}},v\widetilde{\mathbf{D}})$ which by definition satisfy
the conditions (compair with (\ref{lcon})),
\begin{equation}
\widetilde{\mathbf{D}}\widetilde{\theta }=0,\widetilde{\mathcal{Q}}=0%
\mbox{
and }\ h\widetilde{\mathcal{T}}=0,v\widetilde{\mathcal{T}}=0.
\label{cartanlagrc}
\end{equation}%
The geometric d-objects (\ref{dmds}), (\ref{aux21}) and (\ref{dm}) can be
subjected to arbitrary frame transforms on a Lorentz nonholonomic manifold $%
\mathbf{V}$ when we can omit tilde on symbols, for instance, writing such
geometric data in the form $(\mathbf{g,J},\mathbf{P,}),$ but we have to
preserve the notation $\widetilde{\mathbf{D}}$ in all systems of
frames/coordinates because such a d-connection is different, for instance,
from the LC-connection $\nabla .$

We can elaborate on a Lorentz manifold $V$ a relativistic 4-d model of
Finsler space is an example of Lagrange space when a regular $L=F^{2}$ is
defined by a fundamental (generating) Finsler function $F(x,y),$ called also
a Finsler metric, when the nonlinear quadratic element (\ref{nqed}) is
changed into
\begin{equation*}
ds_{F}^{2}=F^{2}(x,y)
\end{equation*}%
and when there are satisfied the conditions: 1) $F$ is a real positive
valued function which is differential on $\widetilde{TV}$ and continuous on
the null section of the projection $\pi :TV\rightarrow V;$ 2) it is imposed
a homogeneity condition $F(x,\lambda y)=|\lambda |$ $F(x,y),$ for a nonzero
real value $\lambda ;$ and 3) the Hessian (\ref{hessls}) is defined by $%
F^{2} $ in such a form that in any point $(x_{(0)},y_{(0)})$ the v-metric is
of signature $(+-).$ The conditions 1-3) allow to construct various types of
geometric models with homogeneity of fiber coordinates with local anisotropy
distinguished on directions. Nevertheless, to extend, for instance, the GR
theory in a relativistic covariant form, we need additional assumptions and
physical motivations on the type of nonlinear and linear connections we
involve into consideration, how to extract effective quadratic elements
etc., see details and references in \cite{vacaru18a,bubuianu18a}. In this
work, we consider that we can always prescribe on a Lorentz manifold $V$ a
Finsler, or Lagrange, type function and state a respective nonholonomic
geometric modeling using canonical data $(\widetilde{L},\ \widetilde{\mathbf{%
N}};\widetilde{\mathbf{e}}_{\alpha },\widetilde{\mathbf{e}}^{\alpha };%
\widetilde{g}_{jk},\widetilde{g}_{ab}),$ when certain homogeneity conditions
can be satisfied for Finsler configurations. For general frame transforms
and modified dispersion relations, we do not consider a Lagrange of Finsler
like nonholonomic variables but can preserve a conventional h- and
v-splitting adapted to a N-connection structure with geometric data $(%
\mathbf{V},\ \mathbf{N};\mathbf{e}_{\alpha },\mathbf{e}^{\alpha
};g_{jk},g_{ab}).$ To elaborate physically realistic gravity models we need
further conventions on the type of linear connection structure (covariant
derivative) we shall use for our geometric constructions.

We can consider always distortion relations
\begin{equation}
\widehat{\mathbf{D}}=\nabla +\widehat{\mathbf{Z}},\widetilde{\mathbf{D}}%
=\nabla +\widetilde{\mathbf{Z}},\mbox{ and }\widehat{\mathbf{D}}=\widetilde{%
\mathbf{D}}+\mathbf{Z,}\mbox{  all determined by }(\mathbf{g,N)\sim }(%
\widetilde{\mathbf{g}}\mathbf{,}\widetilde{\mathbf{N}}\mathbf{)},
\label{candistr}
\end{equation}%
with distortion d-tensors $\widehat{\mathbf{Z}},\widetilde{\mathbf{Z}},$ and
$\mathbf{Z}$, and postulate the (modified) Einstein equations (\ref{cdeinst}%
) in various forms%
\begin{eqnarray}
\widetilde{\mathbf{R}}_{\ \beta \gamma } &=&\widetilde{\Upsilon }_{\ \beta
\gamma }[\widetilde{\mathbf{Z}},\widehat{\mathbf{T}}_{\alpha \beta }],%
\mbox{
or }  \label{lagreq} \\
R_{\ \beta \gamma }[\nabla ] &=&\Upsilon _{\ \beta \gamma }[\widehat{\mathbf{%
Z}},\widehat{\mathbf{T}}_{\alpha \beta }],  \label{standeq}
\end{eqnarray}%
where the (effective) matter sources are respective functionals on
distortions and energy-momentum tensors for matter fields. Such systems of
nonlinear PDEs are different and characterized by different types of Bianchi
identities, local conservation laws and associated symmetries. Nevertheless,
we can establish such classes of nonholonomic frame and distortion
structures, with respective equivalence relations
\begin{equation*}
(\mathbf{g,N,}\widehat{\mathbf{D}})\leftrightarrows (L:\widetilde{\mathbf{g}}%
,\widetilde{\mathbf{N}},\widetilde{\mathbf{D}})\leftrightarrow (\widetilde{%
\theta },\widetilde{\mathbf{P}},\widetilde{\mathbf{J}},\widetilde{\mathbf{D}}%
)\leftrightarrow \lbrack (\mathbf{g},\nabla )]
\end{equation*}%
when the equations (\ref{cdeinst}), (\ref{lagreq}) and (\ref{standeq})
describe equivalent gravitational and matter field models. Different
geometric data have their priorities in constructing in explicit form
different classes of exact/ parametric / approximate solutions or for
performing certain procedures of quantization and further generalizations of
physical theories. If we work with a respective canonical d-connection
structure $\widehat{\mathbf{D}},$ we can prove a general decoupling property
of (\ref{cdeinst}) and construct exact solutions with generic off-diagonal
metrics $g_{\underline{\alpha }\underline{\beta }}(u^{\gamma })$ (\ref{offn}%
) being represented as d-metrics $\mathbf{g}_{\alpha ^{\prime }\beta
^{\prime }}(x,y)$ (\ref{dm}), when the coefficients of such metrics and
associated nonlinear and linear connection structures depend, in principle,
on all spacetime coordinates $u^{\gamma }.$ We can not decouple in general
form the systems of nonlinear PDEs (\ref{lagreq}), in Lagrange-Finsler
variables, and (\ref{standeq}), in local coordinates and for the
LC-connection. In MGTs with modifications of (\ref{lagreq}) or (\ref{standeq}%
), even in GR, we are able to find exact solutions for some "special" ansatz
of metrics which, for instance are diagonalizable and depend only on a
radial or time like coordinate (for instance, for black hole and/or
cosmological solutions). In this work, we shall apply the AFDM in order to
construct cosmological locally anisotropic solutions in MGTs with (in
general, generic off-diagonal) metrics of type $\mathbf{g}_{\alpha ^{\prime
}\beta ^{\prime }}(x^{i},y^{4}=t).$ In geometric and analytic form, this is
possible if we work with nontrivial N-connection structures and certain
variables which are similar to those in Lagrange-Finsler geometry but on
Lorentz manifolds. The almost symplectic Lagrange-Finsler variables $(%
\widetilde{\theta },\widetilde{\mathbf{P}},\widetilde{\mathbf{J}},\widetilde{%
\mathbf{D}})$ have the priority that they allow to elaborate on deformation
quantization and together with $(\mathbf{g,N,}\widehat{\mathbf{D}})$ allow
to introduce nonholonomic and Finsler like spinors and, for instance,
nonholonomic Einstein-Finsler-Dirac systems. This is not possible if the
so-called Berwald- or Chern-Finsler connections are used because they are
not metric compatible and it is a problem to define in a self-consistent
form locally anisotropic versions of the Dirac equation.

\section{TMTs and other MGTs in canonical nonholonomic variables}

\label{mgttmt}

The goal of this section is to show how various classes of MGTs can be
extracted from certain effective Einstein gravity theories using
nonholonomic or Finsler like variables. This allows to decouple the
gravitational field equations and to generate exact solutions in very
general forms, with generic off-diagonal metrics and generalized
connections, and with constraints to zero-torsion configurations, see
details in Refs. \cite%
{vexactsol1,vexactsol1a,vexactsol2,vexactsol3,bubuianu18,bubuianu19,bubuianu20,vacaru20}%
.

In \cite%
{bimmasv1,bimmasv2,vacaru18,vexactsol1,vexactsol1a,vexactsol2,vexactsol3,bubuianu18,bubuianu19,bubuianu20,vacaru20,rajpootvacaru}%
, there were analyzed different possibilities to model different MGTs by
imposing corresponding nonholonomic constraints on the metric and canonical
d--connection structures and source in (\ref{cdeinst}). One of the main
goals of this work is to prove that using corresponding type
parameterizations of the effective Lagrangian $\ \widehat{\mathcal{L}}$ in (%
\ref{estrt}) the so--called modified massive gravity theories (in general,
with bi--connection and bi--metric structures) can be modeled at TMTs with
effective Einstein equations for $\widehat{\mathbf{D}}$ when additional
constraints $\widehat{\mathbf{D}}_{\mid \widehat{\mathcal{T}}=0}=\nabla $
have to be imposed in order to extract LC--configurations.

The actions for equivalent TMT, MGT and nonholonomically deformed Einstein
models are postulated:
\begin{eqnarray}
\mathcal{S} &=&(\ ^{Pl}M)^{2}\int d^{4}u\sqrt{|\widehat{\mathbf{g}}|}[%
\widehat{R}+\widehat{\mathcal{L}}]=  \label{action1} \\
\ ^{\Phi }\mathcal{S}+\ ^{m}\mathcal{S} &=&\int d^{4}u\ ^{1}\Phi (\mathbf{A})%
\left[ \widehat{R}+\ ^{1}L\right] +  \label{action2} \\
&& \int d^{4}u \ ^{2}\Phi (\mathbf{B})\left[ \ ^{2}L+\epsilon \mathbf{f}(%
\widetilde{\mathbf{R}})+(\sqrt{|\mathbf{g}|})^{-1}\ \Phi (\mathbf{H})\right]
+\int d^{4}u\sqrt{|\widehat{\mathbf{g}}|}\ ^{m}\mathcal{L}=  \notag \\
\ ^{F,\mu }\mathcal{S}+\ ^{m}\mathcal{S} &=&(\ ^{Pl}M)^{2}\int d^{4}u[\
\sqrt{|\widehat{\mathbf{g}}|}\ ^{F,\mu }\mathcal{L}+\sqrt{|\widehat{\mathbf{g%
}}|}\ ^{m}\mathcal{L}],  \label{action3}
\end{eqnarray}%
where $|\widehat{\mathbf{g}}|=\det |\widehat{\mathbf{g}}_{\alpha \beta }|$
for a d--metric, $\widehat{\mathbf{g}}_{\alpha \beta }$, constructed
effectively by a conformal transform of a TMT reference one, $\mathbf{g}%
_{\alpha \beta }$, (see below, formula (\ref{confdm}));\ $\ ^{\Phi }L$
defines a class of theories with two independent non--Riemannian
volume--forms $\ ^{1}\Phi (A)$ and $\ ^{2}\Phi (B)$ as in \cite%
{guend3a,guend3b} but with a more general functional for modification, of
type $\epsilon \mathbf{f}(\mathbf{\check{R}}),$ than $\epsilon R^{2}$ if $%
\widehat{\mathbf{D}}\rightarrow \nabla $; the Lagrange density functional $%
^{f,\mu }\mathcal{L}=\mathbf{F}(\widetilde{\mathbf{R}})$ is determined
similar to a modified massive gravity by a mass--deformed scalar curvature
\cite{saridod1,saridod2,saridod3,bimmasv1,bimmasv2},\footnote{%
there are various ambiguities and controversies in different approaches to
massive gravity when modifications by mass terms are postulated for
different Lagrange densities; in this paper, we consider a "toy model" when
terms of type $\mathbf{f}(\mathbf{\check{R}},\mu )$ and/or $\mathbf{f}(%
\mathbf{R})+\mu ...$ can modeled by the same MGT but for different classes
of nonholonomic constraints and different classes of solutions}
\begin{equation}
\mathbf{\check{R}}:=\ \widehat{\mathbf{R}}+2~\mu ^{2}(3-tr\sqrt{\mathbf{g}%
^{-1}\mathbf{q}}-\det \sqrt{\mathbf{g}^{-1}\mathbf{q}}),  \label{auxsc}
\end{equation}%
where $\mu $ is the graviton's mass and $\mathbf{q}=\{\mathbf{q}_{\alpha
\beta }\}$ is the so--called non--dynamical reference metric; $\ ^{m}%
\mathcal{L}$ is the Lagrangian for matter fields.

\subsection{Nonholonomic ghost--free massive configurations}

The term $\epsilon \mathbf{f}(\mathbf{\check{R}})$ in (\ref{action2})
contains possible contributions from a nontrivial graviton mass. Such a
theory can be constructed to be ghost free for very special conditions \cite%
{bimmasv1,bimmasv2}, see explicit results and discussions on possible
applications in modern cosmology in Refs. \cite{saridod1,saridod2,saridod3}.
In this section, we show how prescribing necessary type nonholonomic
configurations such a theory can be equivalently realized as a TMT one
(taking equal actions (\ref{action2}) and (\ref{action3})). For any $(%
\widehat{\mathbf{g}},\mathbf{N},\widehat{\mathbf{D}}),$ we consider the
d--tensor $(\sqrt{\widehat{\mathbf{g}}^{-1}\mathbf{q}})_{~\nu }^{\mu }$
computed as the square root of $\widehat{\mathbf{g}}^{\mu \rho }\mathbf{q}%
_{\rho \nu },$ where
\begin{equation*}
(\sqrt{\widehat{\mathbf{g}}^{-1}\mathbf{q}})_{~\rho }^{\mu }(\sqrt{\widehat{%
\mathbf{g}}^{-1}\mathbf{q}})_{~\nu }^{\rho }=\widehat{\mathbf{g}}^{\mu \rho }%
\mathbf{q}_{\rho \nu },\mbox{ and }\ \sum\limits_{k=0}^{4}~^{k}\beta ~e_{k}(%
\sqrt{\widehat{\mathbf{g}}^{-1}\mathbf{q}})=3-tr\sqrt{\widehat{\mathbf{g}}%
^{-1}\mathbf{q}}-\det \sqrt{\widehat{\mathbf{g}}^{-1}\mathbf{q}},
\end{equation*}%
for some coefficients $~^{k}\beta .$ The values $e_{k}(\mathbf{Y})$ are
defined for a d--tensor $\mathbf{Y}_{~\rho }^{\mu }$ and $Y=[Y]:=tr(\mathbf{Y%
})=\mathbf{Y}_{~\mu }^{\mu },$ where
\begin{eqnarray*}
e_{0}(Y) &=&1,e_{1}(Y)=Y,2e_{2}(Y)=Y^{2}-[Y^{2}],\
6e_{3}(Y)=Y^{3}-3Y[Y^{2}]+2[Y^{3}], \\
24e_{4}(Y) &=&Y^{4}-6Y^{2}[Y^{2}]+3[Y^{2}]^{2}+8Y[Y^{3}]-6[Y^{4}];\
e_{k}(Y)=0\mbox{ for }k>4.
\end{eqnarray*}%
We chose the functional for Lagrange density in (\ref{action3}) in the form $%
\ ^{F,\mu }\mathcal{L}=\mathbf{F}(\mathbf{\tilde{R}}),$ where the functional
dependence $\mathbf{F}$ is different (in general) from $\mathbf{f}(%
\widetilde{\mathbf{R}}).$ For simplicity, we consider Lagrange densities for
matter,$~^{m}\mathcal{L},$ which only depend on the coefficients of a metric
field and not on their derivatives. The energy--momentum d--tensor can be
computed via N--adapted variational calculus,
\begin{equation}
\ ^{m}\widehat{\mathbf{T}}_{\alpha \beta }:=-\frac{2}{\sqrt{|\widehat{%
\mathbf{g}}_{\mu \nu }|}}\frac{\delta (\sqrt{|\widehat{\mathbf{g}}_{\mu \nu
}|}\ \ ^{m}\mathcal{L})}{\delta \widehat{\mathbf{g}}^{\alpha \beta }}=\ ^{m}%
\mathcal{L}\widehat{\mathbf{g}}^{\alpha \beta }+2\frac{\delta (\ ^{m}%
\mathcal{L})}{\delta \widehat{\mathbf{g}}_{\alpha \beta }}.  \label{ematter}
\end{equation}%
Applying such a calculus to $\ ^{F,\mu }\mathcal{S+\ }^{m}\mathcal{S},$ with
$\ ^{1}\mathbf{F}(\mathbf{\check{R}}):=d\mathbf{F}(\mathbf{\check{R}})/d%
\mathbf{\check{R},}$ see details in \cite%
{saridod1,saridod2,saridod3,bimmasv1,bimmasv2}, we obtain the modified
gravitational field equations
\begin{equation}
\widehat{\mathbf{R}}_{\mu \nu }=\ ^{F,\mu }\widehat{\mathbf{\Upsilon }}_{\mu
\nu },  \label{mfeq}
\end{equation}%
where $\ ^{F,\mu }\widehat{\mathbf{\Upsilon }}_{\mu \nu }=~^{m}\widehat{%
\mathbf{\Upsilon }}_{\mu \nu }+~^{f}\widehat{\mathbf{\Upsilon }}_{\mu \nu
}+~^{\mu }\widehat{\mathbf{\Upsilon }}_{\mu \nu },$ for{\small
\begin{eqnarray}
~^{m}\widehat{\mathbf{\Upsilon }}_{\mu \nu } &=&\frac{1}{2M_{P}^{2}}\ ^{m}%
\widehat{\mathbf{T}}_{\alpha \beta },\ \ ^{f}\widehat{\mathbf{\Upsilon }}%
_{\mu \nu }=(\frac{\mathbf{F}}{2~^{1}\mathbf{F}}-\frac{\widehat{\mathbf{D}}%
^{2}\ ^{1}\mathbf{F}}{~^{1}\mathbf{F}})\widehat{\mathbf{g}}_{\mu \nu }+\frac{%
\widehat{\mathbf{D}}_{\mu }\widehat{\mathbf{D}}_{\nu }\ ^{1}\mathbf{F}}{~^{1}%
\mathbf{F}},  \label{source} \\
~^{\mu }\widehat{\mathbf{\Upsilon }}_{\mu \nu } &=&-2\mu ^{2}[(3-tr\sqrt{%
\widehat{\mathbf{g}}^{-1}\mathbf{q}}-\det \sqrt{\widehat{\mathbf{g}}^{-1}%
\mathbf{q}})-\frac{1}{2}\det \sqrt{\widehat{\mathbf{g}}^{-1}\mathbf{q}})]%
\widehat{\mathbf{g}}_{\mu \nu }+  \notag \\
&& \frac{\mu ^{2}}{2}\{\mathbf{q}_{\mu \rho }[(\sqrt{\widehat{\mathbf{g}}%
^{-1}\mathbf{q}})^{-1}]_{~\nu }^{\rho }+\mathbf{q}_{\nu \rho }[(\sqrt{%
\widehat{\mathbf{g}}^{-1}\mathbf{q}})^{-1}]_{~\mu }^{\rho }\}.  \notag
\end{eqnarray}%
} The field equations for massive gravity (\ref{mfeq}) are constructed as
nonholonomic deformations of the Einstein equations (\ref{cdeinst}) when the
source $\widehat{\Upsilon }_{\ \beta \gamma }\rightarrow \ ^{F,\mu }\widehat{%
\mathbf{\Upsilon }}_{\mu \nu }.$

\subsection{TMT massive configurations with (broken) global scaling
invariance}

Let us explain the notations and terms used in above actions chosen in such
forms that a TMT (\ref{action2}) is equivalent to a massive MGT model (\ref%
{action3}) when both classes of such theories are encoded via corresponding
nonholonomic structures into a nonholonomically deformed Einstein gravity
model (\ref{action1}). The non--Riemannian volume--forms (integration
measures on nonholonomic manifold $(\mathbf{g},\mathbf{N},\widehat{\mathbf{D}%
})$) in (\ref{action2}) are determined by two auxiliary 3--index
antisymmetric d--tensor fields $\mathbf{A}_{\alpha \beta \gamma }$ and $%
\mathbf{B}_{\alpha \beta \gamma },$ when
\begin{equation*}
\mathcal{\ }^{1}\Phi (\mathbf{A}):=\frac{1}{3!}\varepsilon ^{\mu \alpha
\beta \gamma }\mathbf{e}_{\mu }\mathbf{A}_{\alpha \beta \gamma }\mbox{ and }%
\mathcal{\ }^{2}\Phi (\mathbf{B}):=\frac{1}{3!}\varepsilon ^{\mu \alpha
\beta \gamma }\mathbf{e}_{\mu }\mathbf{B}_{\alpha \beta \gamma }.
\end{equation*}%
Nevertheless, for non--triviality of the TMT model is crucial the presence
of the 3d auxiliary antisymmetric d--tensor gauge field $\mathbf{H}_{\alpha
\beta \gamma },$ when $\mathcal{\ }\Phi (\mathbf{H}):=\frac{1}{3!}%
\varepsilon ^{\mu \alpha \beta \gamma }\mathbf{e}_{\mu }\mathbf{H}_{\alpha
\beta \gamma }.$ In order to model in some limits two flat regions for the
inflationary and accelerating universe, we consider two Lagrange densities
for a scalar field%
\begin{eqnarray}
\ ^{1}L &=&-\frac{1}{2}\mathbf{g}^{\mu \rho }(\mathbf{e}_{\mu }\varphi )(%
\mathbf{e}_{\rho }\varphi )-\ ^{1}U(\varphi ),\ \ ^{1}U(\varphi )=\
^{1}ae^{-q\varphi };  \label{scalarlagr} \\
\ ^{2}L &=&-\frac{\ ^{2}b}{2}e^{-q\varphi }\mathbf{g}^{\mu \rho }(\mathbf{e}%
_{\mu }\varphi )(\mathbf{e}_{\rho }\varphi )+\ ^{2}U(\varphi ),\ \
^{2}U(\varphi )=\ ^{2}ae^{-2q\varphi },  \notag
\end{eqnarray}%
with dimensional positive parameters $q,\ ^{1}a,\ ^{2}a$ and a dimensionless
one $\ ^{2}b.$ The action (\ref{action2}) is invariant under global
N--adapted Weyl--scale transforms with a positive scale parameter $\lambda ,$
$\mathbf{g}_{\alpha \beta }\rightarrow \lambda \mathbf{g}_{\alpha \beta
},\varphi \rightarrow \varphi +q^{-1}\ln \lambda ,$ $\mathbf{A}_{\alpha
\beta \gamma }\rightarrow \lambda \mathbf{A}_{\alpha \beta \gamma },$ $%
\mathbf{B}_{\alpha \beta \gamma }\rightarrow \lambda ^{2}\mathbf{B}_{\alpha
\beta \gamma }$ and $\mathbf{H}_{\alpha \beta \gamma }\rightarrow \mathbf{H}%
_{\alpha \beta \gamma }.$ For holonomic configurations and quadratic
functionals on LC--scalar $\mathbf{f}(\mathbf{\check{R}})\rightarrow R^{2},$
such a theory is equivalent to that elaborated in \cite%
{guend1a,guend1b,guend1c,guend3a,guend3b}. In a more general context, the
developments in this work involve non-quadratic nonlinear and nonholonomic
functionals and mass gravity deformations via $\mathbf{\check{R}}$ (\ref%
{auxsc}) and generic off--diagonal interactions encoded in $\widehat{\mathbf{%
R}}.$

A variational N--adapted calculus on form fields $\mathbf{A,B,H}$ and on
d--metric $\mathbf{g}$ (with respect to coordinate bases and for $\nabla $
being similar to that presented in section 2 of \cite{guend3a,guend3b})
results in effective gravitational field equations%
\begin{equation}
\widehat{\mathbf{R}}_{\mu \nu }[\widehat{\mathbf{g}}_{\alpha \beta }]=\ ^{ef}%
\widehat{\mathbf{\Upsilon }}_{\mu \nu }+\ ^{F,\mu }\widehat{\mathbf{\Upsilon
}}_{\mu \nu },  \label{densttmt}
\end{equation}%
where $\ ^{F,\mu }\widehat{\mathbf{\Upsilon }}_{\mu \nu }$ is determined by (%
\ref{source}) and $\ ^{ef}\widehat{\Upsilon }_{\ \beta \gamma }:=\varkappa
(\ ^{ef}\widehat{\mathbf{T}}_{\alpha \beta }-\frac{1}{2}\widehat{\mathbf{g}}%
_{\alpha \beta }\ ^{ef}\widehat{T})$ is computed using formulas (\ref{estrt}%
) and (\ref{ematter}) for $\ \mathbf{g}_{\alpha \beta }\rightarrow \widehat{%
\mathbf{g}}_{\alpha \beta }$ and $\ \widehat{\mathcal{L}}\rightarrow \ ^{ef}%
\mathcal{L},$ where%
\begin{eqnarray}
\widehat{\mathbf{g}}_{\alpha \beta } &=&\Theta \mathbf{g}_{\alpha \beta },%
\mbox{ for }\Theta =\ ^{1}\chi -\ ^{2}\chi \epsilon \ ^{1}\mathbf{f}(\
^{1}L+\ ^{1}M,\mu );  \label{confdm} \\
\ ^{ef}L &=&\Theta ^{-1}\left\{ \ ^{1}L+\ ^{1}M+\ ^{2}\chi \Theta ^{-1}\left[
\ ^{2}L+\ ^{1}M+\epsilon \ ^{1}\mathbf{f}(\ ^{1}L+\ ^{1}M,\mu )\right]
\right\} ,  \notag
\end{eqnarray}%
when the conformal factor $\Theta $ for the Weyl re-scaling of d--metric is
induced by the nonlinear functional in the action
\begin{equation}
\ ^{1}\mathbf{f}(\ ^{1}L+\ ^{1}M,\mu )=\frac{d\mathbf{f}(\widehat{\mathbf{R}}%
,\mu )}{d\widehat{\mathbf{R}}}\mid _{\widehat{\mathbf{R}}=\ ^{1}L+\ ^{1}M}
\label{starobrelation}
\end{equation}%
and the two measure functionals $\ ^{1}\chi =\mathcal{\ }^{1}\Phi (\mathbf{A}%
)/\sqrt{|\widehat{\mathbf{g}}_{\mu \nu }|}$ and $\ ^{2}\chi =\mathcal{\ }%
^{2}\Phi (\mathbf{B})/\sqrt{|\widehat{\mathbf{g}}_{\mu \nu }|}.$

The variations on auxiliary anti--symmetric form fields impose certain
constants
\begin{equation*}
\mathbf{e}_{\mu }(\widehat{R}+\ ^{1}L)=0,\mathbf{e}_{\mu }[\ ^{2}L+\epsilon
\mathbf{f}(\mathbf{\check{R}})+\mathcal{\ }\Phi (\mathbf{H})/\sqrt{|\mathbf{g%
}|}]=0,\mathbf{e}_{\mu }[\mathcal{\ }^{2}\Phi (\mathbf{B})/\sqrt{|\widehat{%
\mathbf{g}}_{\mu \nu }|}]=0.
\end{equation*}%
The nonconstant solutions of such nonholonomic constraints allow to preserve
the global Weyl--scale invariance for certain configurations. If we take
constant values
\begin{equation}
\widehat{R}+\ ^{1}L=-\ ^{1}M=const\mbox{ and }\ ^{2}L+\epsilon \mathbf{f}(%
\mathbf{\check{R}})+\mathcal{\ }\Phi (\mathbf{H})/\sqrt{|\mathbf{g}|}=-\
^{2}M=const  \label{constcond}
\end{equation}%
we select configuration with nonholonomic dynamical spontaneous breakdown of
global Weyl--scale invariance when the condition
\begin{equation}
\mathcal{\ }^{2}\Phi (\mathbf{B})/\sqrt{|\widehat{\mathbf{g}}|}=\ ^{2}\chi
=const  \label{ksiconst}
\end{equation}%
preserves the scale invariance. There are certain constraints on the scale
factor $\ ^{1}\chi =\mathcal{\ }^{1}\Phi (\mathbf{A})/\sqrt{|\widehat{%
\mathbf{g}}|}$, which can be derived from variation of (\ref{action2}) on $%
\mathbf{g}_{\mu \nu }$ in N--adapted form. The conditions (\ref{constcond})
relate $\ ^{1}\chi $ and $\ ^{2}\chi ,$ i.e. the integration measures, to
traces $\ ^{1,2}\mathbf{T}:=\mathbf{g}^{\alpha \beta }$ $\ ^{1,2}\mathbf{T}%
_{\alpha \beta }$ of the energy momentum tensors
\begin{equation*}
\ ^{1,2}\mathbf{T}_{\alpha \beta }=\mathbf{g}_{\alpha \beta }\
^{1,2}L-2\partial (\ ^{1,2}L)/\partial \mathbf{g}^{\alpha \beta }
\end{equation*}
of Lagrangians for scalar fields (\ref{scalarlagr}).\footnote{%
For simplicity, we consider matter actions which only depend on the
coefficients of certain effective metric fields and not on their derivatives.%
} This follows form the N--adapted variation on $\mathbf{g}_{\alpha \beta }$
of the action (\ref{action2}) taken for simplicity with $\ $zero $\ ^{m}%
\mathcal{L}.$ which results in%
\begin{equation}
2\ ^{1}\chi \left[ \widehat{\mathbf{R}}_{\mu \nu }(\mathbf{g}_{\alpha \beta
})+\mathbf{g}_{\mu \nu }\ ^{1}L-\ ^{1}\mathbf{T}_{\mu \nu }\right] -\
^{2}\chi \left[ \ ^{2}\mathbf{T}_{\mu \nu }+\mathbf{g}_{\mu \nu }\left(
\epsilon \mathbf{f}(\widetilde{\mathbf{R}})+\ ^{2}M\right) -\ ^{1}\mathbf{f\
}\widehat{\mathbf{R}}_{\mu \nu }(\mathbf{g}_{\alpha \beta })\right] =0.
\label{2mesfeq}
\end{equation}%
Taking the trace of these equations and using (\ref{constcond}), we obtain
the formula $\ ^{1}\chi =\ ^{2}\chi \frac{\ ^{2}\mathbf{T}+2\ ^{2}M}{2\
^{1}L-\ ^{2}\mathbf{T}-2\ ^{1}M}$, which does not depend on the type of $f$%
--modifications containing possible $\mu $--terms. We conclude that above
considered non--Riemannian integration measures and the interactions of
scalar fields (\ref{scalarlagr}) can be modelled as additional distributions
on nonholonomic manifold $(\mathbf{g},\mathbf{N},\widehat{\mathbf{D}}).$
They modify the conformal factor $\Theta $ (\ref{confdm}) and allows to
express the field equations (\ref{2mesfeq}) in Einstein like form (\ref%
{densttmt}), where $\ ^{F,\mu }\widehat{\mathbf{\Upsilon }}_{\mu \nu }$ is
added as an additional effective matter contribution the source of scalar
fields $\ ^{1,2}\mathbf{T}_{\alpha \beta }.$

It should be noted that using the canonical d--connection we obtain $%
\widehat{\mathbf{D}}_{\alpha }\mathbf{T}^{\alpha \beta }=\mathbf{Q}^{\beta
}\neq 0$, when $\mathbf{Q}_{\beta }[\mathbf{g,N}]$ is completely defined by
the d--metric and chosen N--connection structure. Considering nonholonomic
distortions with $\nabla =\widehat{\mathbf{D}}-\widehat{\mathbf{Z}}$, we
obtain standard relations
\begin{equation*}
\nabla ^{\alpha }(R_{\alpha \beta }-\frac{1}{2}g_{\alpha \beta }R)=0%
\mbox{
and }\nabla ^{\alpha }\Upsilon _{\alpha \beta }=0.
\end{equation*}
A similar property exists in Lagrange mechanics with non--integrable
constraints when the standard conservation laws do not hold true. A new
class of effective variables and new types of conservation laws can be
introduced and, respectively, constructed using Lagrange multiples.

The main conclusion of this section is that various MGTs with two
integration measures, possible deformations by mass graviton terms,
bi--connection and bi--metric structures can be expressed as nonholonomic
deformations of the Einstein equations in the form (\ref{cdeinst}).
Different theories are characterized by respective sources (in explicit
form, $\ ^{F,\mu }\widehat{\mathbf{\Upsilon }}_{\mu \nu }$ in (\ref{mfeq}),
or $\ ^{ef}\widehat{\mathbf{\Upsilon }}_{\mu \nu }+\ ^{F,\mu }\widehat{%
\mathbf{\Upsilon }}_{\mu \nu }$ in (\ref{densttmt})). Our next goal is to
prove that such effective Einstein equations can be integrated in certain
general forms for $\widehat{\mathbf{D}}$ and possible constraints (\ref%
{lccond}) for LC--configurations.

\section{Cosmological Solutions in Effective Einstein Gravity and FMGTs}

\label{scosmsol} We can generate in explicit form integral varieties of
systems of PDEs of type (\ref{cdeinst}) for d--metrics $\widehat{\mathbf{g}}$
(\ref{confdm}) and sources $\widehat{\Upsilon }_{\ \beta \gamma }=\ ^{ef}%
\widehat{\mathbf{\Upsilon }}_{\mu \nu }+\ ^{F,\mu }\widehat{\mathbf{\Upsilon
}}_{\mu \nu }$ as in (\ref{densttmt}) which via frame and coordinate
transforms,
\begin{equation*}
\widehat{\mathbf{g}}_{\alpha \beta }=e_{\ \alpha }^{\alpha ^{\prime
}}(x^{i},y^{a})e_{\ \beta }^{\beta ^{\prime }}(x^{i},y^{a})\widehat{\mathbf{g%
}}_{\alpha ^{\prime }\beta ^{\prime }}(x^{i},t)\mbox{ and }\widehat{\Upsilon
}_{\alpha \beta }=e_{\ \alpha }^{\alpha ^{\prime }}(x^{i},y^{a})e_{\ \beta
}^{\beta ^{\prime }}(x^{i},y^{a})\widehat{\Upsilon }_{\alpha ^{\prime }\beta
^{\prime }}(x^{i},t),
\end{equation*}%
for a time like coordinate $y^{4}=t$ ($i^{\prime },i,k,k^{\prime },...=1,2$
and $a,a^{\prime },b,b^{\prime },...=3,4$), can be parameterized in the form:%
\begin{eqnarray}
\widehat{\mathbf{g}} &=&\widehat{\mathbf{g}}_{\alpha ^{\prime }\beta
^{\prime }}\mathbf{e}^{\alpha ^{\prime }}\otimes \mathbf{e}^{\beta ^{\prime
}}=g_{i}(x^{k})dx^{i}\otimes dx^{j}+\omega ^{2}(x^{k},y^{3},t)h_{a}(x^{k},t)%
\mathbf{e}^{a}\otimes \mathbf{e}^{a},  \label{dm1} \\
\mathbf{e}^{3} &=&dy^{3}+n_{i}(x^{k},t)dx^{i},\mathbf{e}%
^{4}=dt+w_{i}(x^{k},t)dx^{i},  \notag
\end{eqnarray}%
for nontrivial
\begin{eqnarray*}
&&\{g_{i^{\prime }j^{\prime }}\}=diag[g_{i}],g_{1}=g_{2}=e^{\psi
(x^{k})};\{g_{a^{\prime }b^{\prime }} \}= diag[h_{a}],h_{a}=h_{a}(x^{k},t);
\\
&&N_{i}^{3}=n_{i}(x^{k},t);\ N_{i}^{4}=w_{i}(x^{k},t);
\end{eqnarray*}
and%
\begin{eqnarray}
\widehat{\Upsilon }_{\alpha ^{\prime }\beta ^{\prime }} &=&diag[\Upsilon
_{i};\Upsilon _{a}],\mbox{ for }\Upsilon _{1}=\Upsilon _{2}=\widetilde{%
\Upsilon }(x^{k})=\ ^{ef}\widetilde{\Upsilon }(x^{k})+~^{m}\widetilde{%
\Upsilon }(x^{k})+~^{f}\widetilde{\Upsilon }(x^{k})+~^{\mu }\widetilde{%
\Upsilon }(x^{k}),  \notag \\
\Upsilon _{3} &=&\Upsilon _{4}=\Upsilon (x^{k},t)=\ ^{ef}\Upsilon
(x^{k},t)+~^{m}\Upsilon (x^{k},t)+~^{f}\Upsilon (x^{k},t)+~^{\mu }\Upsilon
(x^{k},t).  \label{2sourc}
\end{eqnarray}
These ansatz for the d--metric and sources are very general one but for an
assumption that there are N--adapted frames with respect to which the MGTs
interactions are with Killing symmetry on $\partial /\partial y^{3}$ when
geometric and physical values do not depend on coordinate $y^{3}.$\footnote{%
It should be noted that it is possible to construct very general classes of
generic off--diagonal solutions depending on all spacetime variables in
arbitrary finite dimensions, see details and examples in Refs. \cite%
{vexactsol1,vexactsol1a,vexactsol2,vexactsol3} for more "non--Killing"
configurations. For simplicity, we shall study in this work nonhomogeneous
and locally anisotropic cosmological solutions depending on variables $%
(x^{k},t)$ with smooth limits to cosmological diagonal configurations
depending only on $t$ and very small off--diagonal contributions
characterized by a small parameter $\varepsilon ,$ $0\leq \varepsilon \ll 1$%
).} We use parameterizations $g_{1}=g_{2}=e^{\psi (x^{i})}$ and $%
h_{a}(x^{k},t)$ for $\ i,j,...=1,2$ and $a,b,...=3,4;$ and N--connection
coefficients \ $\mathbf{N}_{i}^{3}=n_{i}(x^{k},t)$ and $\mathbf{N}%
_{i}^{4}=w_{i}(x^{k},t).$ Introducing brief denotations for partial
derivatives like $a^{\bullet }=\partial _{1}a,b^{\prime }=\partial
_{2}b,h^{\ast }=\partial _{4}h=\partial _{t}h$ $\ $and defining the values $%
\alpha _{i}=h_{3}^{\ast }\partial _{i}\varpi ,\beta =h_{3}^{\ast }\ \varpi
^{\ast },\gamma =\left( \ln |h_{3}|^{3/2}/|h_{4}|\right) ^{\ast }$ for a
generating function
\begin{equation}
\varpi :=\ln |h_{3}^{\ast }/\sqrt{|h_{3}h_{4}|}|,%
\mbox{ we shall also use
the value  }\Psi :=e^{\varpi },  \label{genf}
\end{equation}%
we transform (\ref{densttmt}) into a nonlinear system of PDEs with
decoupling property for the un--known functions $\psi
(x^{i}),h_{a}(x^{k},t),w_{i}(x^{k},t)$ and $n_{i}(x^{k},t),$
\begin{equation}
\psi ^{\bullet \bullet }+\psi ^{\prime \prime }=2~\widetilde{\Upsilon },\
\varpi ^{\ast }h_{3}^{\ast }=2h_{3}h_{4}\Upsilon ,\ n_{i}^{\ast \ast
}+\gamma n_{i}^{\ast }=0,\ \beta w_{i}-\alpha _{i}=0.\   \label{eq4}
\end{equation}%
This system posses another very important property which allows us to
re--define the generating function, $\Psi \longleftrightarrow \widetilde{%
\Psi },$ when $\Lambda (\Psi ^{2})^{\ast }=|\Upsilon |(\widetilde{\Psi }%
^{2})^{\ast }$ and
\begin{equation}
\Lambda \Psi ^{2}=\widetilde{\Psi }^{2}|\Upsilon |+\int dt\widetilde{\Psi }%
^{2}|\Upsilon |^{\ast }  \label{nonltr}
\end{equation}%
for $\widetilde{\Psi }:=\exp \widetilde{\varpi }$ and any prescribed values
of effective (for different types of contributions $ef,m,f,\mu $)
cosmological constants in $\Lambda =\ ^{ef}\Lambda +~^{m}\Lambda
+~^{f}\Lambda +~^{\mu }\Lambda $ associated respectively to%
\begin{equation*}
\Upsilon (x^{k},t)=\ ^{ef}\Upsilon (x^{k},t)+~^{m}\Upsilon
(x^{k},t)+~^{f}\Upsilon (x^{k},t)+~^{\mu }\Upsilon (x^{k},t).
\end{equation*}
For generating off--diagonal cosmological solutions depending on $t,$ we
have to consider generating functions for which $\Psi ^{\ast }\neq 0.$ The
equations (\ref{eq4}) for ansatz (\ref{dm1}) transform respectively into
such a system of nonlinear PDEs%
\begin{eqnarray}
&&\psi ^{\bullet \bullet }+\psi ^{\prime }=2~\widetilde{\Upsilon },\
\widetilde{\varpi }^{\ast }h_{3}^{\ast }=2h_{3}h_{4}\Lambda ,\ n_{i}^{\ast
\ast }+\gamma n_{i}^{\ast }=0,\ \ \varpi ^{\ast }w_{i}-\partial _{i}\varpi =0
\label{eq4a} \\
&&\mbox{and }\varpi ^{\ast }\partial _{i}\omega -\omega ^{\ast }\partial
_{i}\varpi =0,\mbox{ for the vertical conformal factor}.  \notag
\end{eqnarray}%
We have to subject the d--metric and N--connection coefficients to
additional constraints (\ref{lccond}) in order to satisfy the torsionless
conditions, which for the ansatz (\ref{dm1}) are written
\begin{equation}
w_{i}^{\ast }=(\partial _{i}-w_{i}\partial _{4})\ln \sqrt{|h_{4}|},(\partial
_{i}-w_{i}\partial _{4})\ln \sqrt{|h_{3}|}=0,\partial _{i}w_{j}=\partial
_{j}w_{i},n_{i}^{\ast }=0,\partial _{i}n_{j}=\partial _{j}n_{i}.
\label{lccond1}
\end{equation}

We can generate exact solutions in TMT, MGT and nonholonomically deformed
Einstein theories with respective actions (\ref{action1}), (\ref{action2})
and (\ref{action3}) using integral varieties\footnote{%
The term "integral variety" is used in the theory of differential equations
for certain "classes of solutions" determined by corresponding classes of
parameters, generating and integration functions etc. In GR, we search for
an integral variety of solutions of associated systems of PDEs determining,
for instance, Einstein spacetimes, black holes, cosmological solutions etc.
In modified gravity theories, we can keep certain analogy to GR if consider
effective models.} of the system of PDEs (\ref{eq4a}) which can be found in
very general forms. Let us briefly explain this geometric formalism
elaborated in the framework of the AFDM (see details, for instance, in Refs.
\cite{bubuianu18,bubuianu19,bubuianu20,vacaru20}):

\begin{enumerate}
\item The first equation for $\psi $ is just the 2--d Laplace/ d' Alambert
equation which can be solved for any given $~\widetilde{\Upsilon },$ which
allows us to find $g_{1}=g_{2}=e^{\psi (x^{k})}.$

\item Using the second equation in (\ref{eq4a}) and (\ref{genf}), the
coefficients $h_{a}$ can be expressed as functionals on $(\Psi ,\Upsilon ).$
We re--define the generating function as in (\ref{nonltr}) and consider an
effective source%
\begin{equation*}
\Xi :=\int dt\Upsilon (\widetilde{\Psi }^{2})^{\ast }=\ ^{ef}\Xi +\ ^{m}\Xi
+\ ^{f}\Xi +\ ^{\mu }\Xi ,
\end{equation*}%
when $\ ^{ef}\Xi :=\int dt\ ^{ef}\Upsilon (\widetilde{\Psi }^{2})^{\ast },$ $%
\ ^{m}\Xi :=\int dt\ \ ^{m}\Upsilon (\widetilde{\Psi }^{2})^{\ast },$ $\
^{f}\Xi :=\int dt\ \ ^{f}\Upsilon (\widetilde{\Psi }^{2})^{\ast },$ and
write
\begin{equation*}
h_{3}=\frac{\widetilde{\Psi }^{2}}{4(\ ^{ef}\Lambda +~^{m}\Lambda
+~^{f}\Lambda +~^{\mu }\Lambda )}\mbox{ and }h_{4}=\frac{(\widetilde{\Psi }%
^{\ast })^{2}}{(\ ^{ef}\Xi +\ ^{m}\Xi +\ ^{f}\Xi +\ ^{\mu }\Xi )}.
\end{equation*}

\item We have to integrate two times on $t$ in order to find from the 3d
subset of equations in (\ref{eq4a})
\begin{equation*}
n_{i}=\ _{1}n_{i}+\ _{2}n_{i}\int dt\frac{(\widetilde{\Psi }^{\ast })^{2}}{%
\widetilde{\Psi }^{3}(\ ^{ef}\Xi +\ ^{m}\Xi +\ ^{f}\Xi +\ ^{\mu }\Xi )}
\end{equation*}%
for some integration functions $\ _{1}n_{i}(x^{k})$ and $\ _{2}n_{i}(x^{k}).$

\item The 4th set of equations in (\ref{eq4a}) are algebraic ones which
allows us to compute%
\begin{equation*}
w_{i}=[\varpi ^{\ast }]^{-1}\partial _{i}\varpi =[\Psi ^{\ast
}]^{-1}\partial _{i}\Psi =[(\Psi ^{2})^{\ast }]^{-1}\partial _{i}(\Psi
)^{2}=[\Xi ^{\ast }]^{-1}\partial _{i}\Xi .
\end{equation*}

\item We can satisfy the conditions for $\omega $ in the second line in (\ref%
{eq4a}) if we keep, for simplicity, the Killing symmetry on $\partial _{i}$
and take, for instance, $\omega ^{2}=|h_{4}|^{-1}.$

Different types of inhomogeneous cosmological solutions of the system (\ref%
{densttmt}) are determined by corresponding classes of and effective sources
\begin{eqnarray*}
\mbox{generating functions:} && \psi (x^{k}),\widetilde{\Psi }%
(x^{k},t),\omega (x^{k},y^{3},t) \\
\mbox{effective sources: } &&\widetilde{\Upsilon }(x^{k});\ ^{ef}\Xi
(x^{k},t),\ ^{m}\Xi (x^{k},t),\ ^{f}\Xi (x^{k},t),\ ^{\mu }\Xi (x^{k},t), \\
&& \mbox{or }\ ^{ef}\Upsilon (x^{k},t),\ ^{m}\Upsilon (x^{k},t),\
^{f}\Upsilon (x^{k},t),\ ^{\mu }\Upsilon (x^{k},t) \\
\mbox{integration cosm. constants:} && \ ^{ef}\Lambda ,~^{m}\Lambda
,~^{f}\Lambda ,~^{\mu }\Lambda \\
\mbox{integration functions:} && \ _{1}n_{i}(x^{k})\mbox{ and }\
_{2}n_{i}(x^{k})
\end{eqnarray*}%
We can generate solutions with any nontrivial $\ ^{ef}\Lambda ,~^{m}\Lambda
,~^{f}\Lambda ,~^{\mu }\Lambda $ even any, or all, effective source $\
^{ef}\Upsilon ,\ ^{m}\Upsilon ,\ ^{f}\Upsilon ,\ ^{\mu }\Upsilon $ can be
zero.
\end{enumerate}

\subsection{Inhomogeneous FTMT and MGT configurations with induced
nonholonomic torsion}

The solutions with coefficients computed above in 1-5 can be parametrized in
a form to describe nonholonomic deformations, $\widehat{\mathbf{g}}_{\alpha
\beta }=\mathbf{e}_{\ \alpha }^{\alpha ^{\prime }}\mathbf{e}_{\ \beta
}^{\beta ^{\prime }}\mathring{g}_{\alpha ^{\prime }\beta ^{\prime }},$ of
the Friedman--Lema\^{\i}tre--Robertson--Walker, FLRW, diagonal quadratic
element\footnote{%
we can consider spherical symmetry coordinates $u^{\alpha ^{\prime
}}=(x^{1^{\prime }}=r,x^{2^{\prime }}=\theta ,y^{3^{\prime }}=\varphi
,y^{4^{\prime }}=t),$ or Cartesian ones, $u^{\alpha ^{\prime
}}=(x^{1^{\prime }}=x,x^{2^{\prime }}=y,y^{3^{\prime }}=z,y^{4^{\prime
}}=t), $ for a scale factor $\mathring{a}(t)$ determining the Hubble
constant $H:=\mathring{a}^{\ast }/\mathring{a}$}
\begin{equation}
d\mathring{s}^{2}=\mathring{g}_{\alpha ^{\prime }\beta ^{\prime }}du^{\alpha
^{\prime }}du^{\beta ^{\prime }}=\mathring{a}^{2}(t)[dr^{2}+r^{2}d\theta
^{2}+r^{2}\sin ^{2}\theta d\varphi ^{2}]-dt^{2}  \label{flrw}
\end{equation}%
into a generic off--diagonal inhomogeneous cosmological metric of type (\ref%
{dm1}) with $g_{i}=\eta _{i}e^{\psi }$ and $h_{a}=\eta _{a}\mathring{g}_{a}$
with effective polarization functions $\eta _{1}=\eta _{2}=a^{-2}e^{\psi
},\eta _{3}=\mathring{a}^{-2}h_{3},\eta _{4}=1$ and $\widehat{h}%
_{3}=h_{3}/a^{2}|h_{4}|,$ when{%
\begin{eqnarray}
ds^{2} &=&a^{2}(x^{k},t)[\eta _{1}(x^{k},t)(dx^{1})^{2}+\eta
_{2}(x^{k},t)(dx^{2})^{2}]  \notag \\
&&+a^{2}(x^{k},t)\widehat{h}_{3}(x^{k},t)[dy^{3}+(\ _{1}n_{i}+\
_{2}n_{i}\int dt\frac{(\widetilde{\Psi }^{\ast })^{2}}{\widetilde{\Psi }%
^{3}(\ ^{ef}\Xi +\ ^{m}\Xi +\ ^{f}\Xi +\ ^{\mu }\Xi )})dx^{i}]^{2}  \notag \\
&&-[dt+\frac{\partial _{i}(\ ^{ef}\Xi +\ ^{m}\Xi +\ ^{f}\Xi +\ ^{\mu }\Xi )}{%
(\ ^{ef}\Xi +\ ^{m}\Xi +\ ^{f}\Xi +\ ^{\mu }\Xi )^{\ast }}dx^{i}]^{2}.
\label{solnonht}
\end{eqnarray}%
The inhomogeneous scaling factor }$a(x^{k},t)$ in (\ref{solnonht}) is
related to the generating function $\widetilde{\Psi }$ via formula%
\begin{equation*}
a^{2}\widehat{h}_{3}=\omega ^{2}h_{3}=\frac{h_{3}}{|h_{4}|}=\frac{\widetilde{%
\Psi }^{2}|\ ^{ef}\Xi +\ ^{m}\Xi +\ ^{f}\Xi +\ ^{\mu }\Xi |}{4(\
^{ef}\Lambda +~^{m}\Lambda +~^{f}\Lambda +~^{\mu }\Lambda )(\widetilde{\Psi }%
^{\ast })^{2}}.
\end{equation*}%
In general, such target metrics $\widehat{\mathbf{g}}_{\alpha \beta
}(x^{k},t)$ determine new classes of cosmological metrics with nontrivial
nonholonomically induced torsion computed for $\widehat{\mathbf{D}}.$ Such
modified spacetimes can not be diagonalized by coordinate transforms if the
anholonomy coefficients $W_{\alpha \beta }^{\gamma }$ are not zero. For
trivial gravitational polarizations, $\eta _{\alpha }=1,$ trivial
N--connection coefficients, $N_{i}^{3}=n_{i}=0$ and $N_{i}^{4}=w_{i}=0,$ and
for $a(x^{k},t)\rightarrow \mathring{a}(t)$ we obtain torsionless FLRW
metrics. We emphasize that one could not be smooth limits $\widehat{\mathbf{g%
}}_{\alpha \beta }\rightarrow \mathring{g}_{\alpha \beta }$ for arbitrary
generating function $\widetilde{\Psi }$ and any nontrivial effective
cosmological constant $\ ^{ef}\Lambda ,~^{m}\Lambda ,~^{f}\Lambda ,$ or $%
~^{\mu }\Lambda ,$ associated to respective mater fields.

We can generate off--diagonal cosmological configurations as "small"
deformations with $\eta _{\alpha }=1+$ $\epsilon _{\alpha },n_{i}=\
^{\epsilon }n_{i}$ and $w_{i}=\ ^{\epsilon }w_{i},$ with $|\epsilon _{\alpha
}|,|\ ^{\epsilon }n_{i}|,|\ ^{\epsilon }w_{i}|\ll 1.$ In particular, we can
study only TMT models if $\ ^{m}\Xi =\ ^{f}\Xi =\ ^{\mu }\Xi =0$ and $%
~^{m}\Lambda =~^{f}\Lambda =~^{\mu }\Lambda =0$ but $\ ^{ef}\Upsilon
(x^{k},t)\neq 0$ and $\ ^{ef}\Lambda \neq 0.$ Off--diagonal cosmological
scenarios in massive and bi--metric gravity with nontrivial $\ ^{\mu }\Xi $
and $~^{\mu }\Lambda $ were studied in our recent works \cite%
{bimmasv1,bimmasv2}. Other classes of MGTs and cosmological models with
off--diagonal configurations when $f$--modified gravity effects are modelled
in GR were studied in \cite%
{vexactsol1,vexactsol1a,vexactsol2,vexactsol3,bubuianu18,bubuianu19,bubuianu20,vacaru20}%
. The goal of section \ref{smassgrfromtmt} is to show how TMT gravity and
cosmological models can be associated to certain nonholonomic off--diagonal
de Sitter configurations with nontrivial $\ ^{ef}\Lambda $ for an effective
Einstein-Lagrange spacetime and such constructions can be generalized to
reproduce MGTs and massive gravity.

\subsection{Extracting Levi--Civita cosmological configurations}

Let us show how we can generate in explicit form solutions of the system (%
\ref{lccond1}) for nonholonomic generic off--diagonal configurations with
zero torsion. We have to consider certain special classes of generating and
integration functions. By straightforward computations we can check that
such conditions are satisfied if we state such conditions for a metric (\ref%
{solnonht}) that
\begin{eqnarray}
\ _{2}n_{i} &=&0\mbox{ and }\ _{1}n_{i}=\partial _{i}n(x^{k}),%
\mbox{ for any
}n(x^{k})  \label{lccond2} \\
\Psi &=&\check{\Psi},\mbox{ for }(\partial _{i}\check{\Psi})^{\ast
}=\partial _{i}(\check{\Psi}^{\ast })\mbox{ and find a function }\check{A}%
(x^{k},t)\mbox{ when }  \notag \\
\partial _{i}w_{j} &=&\partial _{j}w_{i}\mbox{ for }w_{i}=\check{w}%
_{i}=\partial _{i}\check{\Psi}/\check{\Psi}^{\ast }=\partial _{i}\widehat{%
\Xi }/\widehat{\Xi }^{\ast }=\partial _{i}\check{A}  \notag
\end{eqnarray}%
when
\begin{equation*}
\Lambda \check{\Psi}^{2}=\widehat{\Psi }^{2}|\Upsilon |+\int dt\widehat{\Psi
}^{2}|\Upsilon |^{\ast }\mbox{ and }\widehat{\Xi }:=\int dt\Upsilon (%
\widehat{\Psi }^{2})^{\ast }
\end{equation*}
are computed following formulas (\ref{nonltr}) but for $\Psi (\widetilde{%
\Psi })\rightarrow \check{\Psi}(\widehat{\Psi })$ and $\widetilde{\Psi }%
\rightarrow \widehat{\Psi }.$ For certain configurations, we can consider
functional dependencies $\widehat{\Psi }=\widehat{\Psi }(\ln \sqrt{|h_{3}|})$
and invertible functional dependencies $h_{3}[\widehat{\Psi }[\check{\Psi}%
]]. $ In such cases, we take a $h_{3}(x^{k},t)$ as a generating function and
consider necessary type functionals $\check{\Psi}[h_{3}]$ with the property $%
(\partial _{i}\check{\Psi})^{\ast }=\partial _{i}(\check{\Psi}^{\ast })$
which are used for defining $\check{w}_{i}[h_{3}]=\partial _{i}\check{\Psi}/%
\check{\Psi}^{\ast }=\partial _{i}\check{A}[h_{3}].$

Putting together the conditions (\ref{lccond2}), we generate nonhomogeneous
cosmological LC--configurations with quadratic linear elements{%
\begin{eqnarray}
ds^{2} &=&\check{a}^{2}(x^{k},t)[\eta _{1}(x^{k},t)(dx^{1})^{2}+\eta
_{2}(x^{k},t)(dx^{2})^{2}]+\check{a}^{2}(x^{k},t)\widehat{h}%
_{3}(x^{k},t)[dy^{3}+(\partial _{i}n)dx^{i}]^{2}  \notag \\
&& -[dt+(\partial _{i}\check{A})dx^{i}]^{2}  \notag \\
&=&e^{\psi (x^{k})}[(dx^{1})^{2}+(dx^{2})^{2}]+\frac{\widehat{\Psi }^{2}}{%
4(\ ^{ef}\Lambda +~^{m}\Lambda +~^{f}\Lambda +~^{\mu }\Lambda )}%
[dy^{3}+(\partial _{i}n)dx^{i}]^{2}  \notag \\
&& -\frac{(\widehat{\Psi }^{\ast })^{2}}{\ \widehat{\Xi }}[dt+(\partial _{i}%
\check{A})dx^{i}]^{2}.  \label{lcconf}
\end{eqnarray}%
The inhomogeneous scaling factor }$\check{a}(x^{k},t)$ is computed similarly
to (\ref{solnonht}) but using the generating function $\widehat{\Psi },$
\begin{equation*}
\check{a}^{2}\widehat{h}_{3}=\frac{\widehat{\Psi }^{2}|\ \widehat{\Xi }|}{%
4(\ ^{ef}\Lambda +~^{m}\Lambda +~^{f}\Lambda +~^{\mu }\Lambda )(\widehat{%
\Psi }^{\ast })^{2}}\mbox{ for }\check{\Psi}:=e^{\check{\varpi}}.
\end{equation*}%
Having constructed a class of generic off--diagonal solutions (\ref{lcconf}%
), we can impose additional constraints on the generating/integration
functions and constants and source in order to explain certain observational
cosmological data. For instance, we can fix subclasses of functions \ $%
\widehat{\Psi }\rightarrow \widehat{\Psi }(t),(\partial _{i}\check{A}%
)\rightarrow w_{i}(t)$ etc. describing small deformations of a FLRW metric (%
\ref{flrw}) in a nonlinear parametric way re--defined generating functions (%
\ref{nonltr}) and different types of effective sources in TMT, MGT and/or
massive gravity models.

\section{Locally Anisotropic Effective Scalar Potentials and Flat Regions}

\label{stwoflat} We study three examples of off--diagonal cosmological
solutions reproducing the TMT model with two flat regions of the effective
scalar potental studied in Ref. \cite{guend2}, than analyse how massive
gravity can be modelled as a TMT theory and effective GR, and (in the last
subsection) we speculate on non--singular emergent anisotropic universes.
The solutions in this section will be constructed to contain nontrivial
nonholonomically induced torsion as for quadratic elements (\ref{solnonht}).
For certain important limits, LC--configurations of type (\ref{lcconf}) will
be also examined.

\subsection{Off--diagonal interactions and associated TMT models with two
flat regions}

We chose the nontrivial off--diagonal data in (\ref{solnonht}) for $\
~^{m}\Lambda =~^{f}\Lambda =~^{\mu }\Lambda =0$ and $\ ^{m}\Upsilon \ =\
^{f}\Upsilon \ =\ ^{\mu }\Upsilon \ =0$\ resulting in $^{m}\Xi =\ ^{f}\Xi =\
^{\mu }\Xi =0,$ but consider nonzero $\ ^{ef}\Lambda $ and $\ ^{ef}\Upsilon $
\ is taken as a one--Killing configuration not depending on $y^{3}$ in
\begin{equation*}
\ ^{ef}\widehat{\Upsilon }_{\ \beta \gamma }:=\varkappa (\ ^{ef}\widehat{%
\mathbf{T}}_{\alpha \beta }-\frac{1}{2}\widehat{\mathbf{g}}_{\alpha \beta }\
^{ef}\widehat{T})
\end{equation*}
is computed using formula (\ref{estrt}) and (\ref{ematter}) for $\ \mathbf{g}%
_{\alpha \beta }\rightarrow \widehat{\mathbf{g}}_{\alpha \beta }$ and $\
\widehat{\mathcal{L}}\rightarrow \ ^{ef}\mathcal{L}$ for two scalar
densities (\ref{scalarlagr}) as in (\ref{confdm}). We generate solutions of
\ $\widehat{\mathbf{R}}_{\mu \nu }[\widehat{\mathbf{g}}_{\alpha \beta }]=\
^{ef}\widehat{\mathbf{\Upsilon }}_{\mu \nu }$ (in a particular case of (\ref%
{densttmt})) for $\widehat{\mathbf{g}}_{\alpha \beta }(x^{k},t)=\widehat{%
\Theta }(x^{k},t)\mathbf{g}_{\alpha \beta }(x^{k},t)$, parameterized in the
form {%
\begin{eqnarray}
ds^{2} &=&\widehat{\mathbf{g}}_{\alpha \beta }\mathbf{e}^{\alpha }\mathbf{e}%
^{\beta }=\widetilde{a}^{2}(x^{k},t)[\eta _{1}(x^{k},t)(dx^{1})^{2}+\eta
_{2}(x^{k},t)(dx^{2})^{2}]  \label{offdguend} \\
&&+\widetilde{a}^{2}(x^{k},t)\widehat{h}_{3}(x^{k},t)[dy^{3}+(\ _{1}n_{i}+\
_{2}n_{i}\int dt\frac{(\widetilde{\Psi }^{\ast })^{2}}{\widetilde{\Psi }%
^{3}\ ^{ef}\Xi })dx^{i}]^{2}-[dt+\frac{\partial _{i}(\ ^{ef}\Xi )}{(\
^{ef}\Xi )^{\ast }}dx^{i}]^{2}.  \notag
\end{eqnarray}%
The inhomogeneous scaling factor }$\widetilde{a}(x^{k},t)$ is related to the
generating function $\widetilde{\Psi }$ via formula%
\begin{equation}
\widetilde{a}^{2}\widehat{h}_{3}=\omega ^{2}h_{3}=\frac{h_{3}}{|h_{4}|}=%
\frac{\widetilde{\Psi }^{2}|\ ^{ef}\Xi |}{4(\ ^{ef}\Lambda )(\widetilde{\Psi
}^{\ast })^{2}}.  \label{aux1}
\end{equation}

Choosing a function $\widetilde{\Psi }$, we prescribe a corresponding
dependence for $\widehat{\Theta }(x^{k},t)$ and, respectively, $\widetilde{a}%
(x^{k},t)$ as follow from above formulas. Let us speculate on the structure
of $\widehat{\Theta }$ which describe off--diagonal generalizations of the
model given by formulas (18)--(23) in \cite{guend2} in the assumption that
the relation (\ref{starobrelation}) for zero graviton mass and quadratic
Ricci scalar curvature has the limit
\begin{equation*}
\ ^{1}\mathbf{f}(\ ^{1}L+\ ^{1}M,\mu =0)\mathbf{=}\frac{d\mathbf{f}(\widehat{%
\mathbf{R}},\mu =0)}{d\widehat{\mathbf{R}}}\mid _{\widehat{\mathbf{R}}=\
^{1}L+\ ^{1}M}\rightarrow \ \ ^{1}U-\ ^{1}M.
\end{equation*}
In this subsection, we shall consider $\ ^{1}\mathbf{f\approx }\ \ ^{1}U-\
^{1}M$ for a nonhomogeneous $\varphi (x^{k},t)\mathbf{\approx }\varphi (t)$
in order to construct cosmological TMT models with limits to diagonal two
flat regions.

We consider $\widehat{\Theta }$ as a conformal factor $\Theta $ in (\ref%
{confdm}) not depending on $y^{3}$ written in explicit form for an Einstein
N--adapted frame with effective scalar Lagrangian%
\begin{eqnarray}
\ ^{ef}\widehat{L} &=&\widehat{\Theta }^{-1}\{\ ^{1}\widehat{L}+\ ^{1}M+\
^{2}\chi \widehat{\Theta }^{-1}[\ ^{2}\widehat{L}+\ ^{1}M+\epsilon (\ ^{1}%
\mathbf{f)}^{2}]\}  \notag \\
&=&A(\varphi )X+B(\varphi )X^{2}-\ ^{ef}U(\varphi ),  \label{eff2}
\end{eqnarray}%
where we omit cumbersome formulas for $A(\varphi )$ and $B(\varphi )$ in the
second line (see similar ones given by formulas (24) and (25) in \cite%
{guend2}) but present%
\begin{eqnarray}
\ ^{1}\widehat{L} &=&\widehat{\Theta }\text{ }X-\ \ ^{1}U\mbox{ for }\ ^{2}%
\widehat{L}=\frac{\ ^{2}b}{\ ^{1}a}\widehat{\Theta }\text{ }\ \ ^{1}U%
\widehat{X}+\ \ ^{2}U\mbox{ for }\widehat{X}=-\frac{1}{2}\widehat{\mathbf{g}}%
^{\alpha \beta }\mathbf{e}_{\alpha }\varphi \mathbf{e}_{\beta }\varphi ,
\notag \\
\ ^{ef}U &=&\frac{(\ ^{1}\mathbf{f)}^{2}}{4\ ^{2}\chi \lbrack \ ^{2}U+\
^{2}M+\epsilon (\ ^{1}\mathbf{f)}^{2}]}.  \label{effpot}
\end{eqnarray}%
For simplicity, we can construct off--diagonal configurations with $\widehat{%
h}_{3}\simeq 1$ in (\ref{aux1}), prescribing a value $\ ^{ef}\Lambda $
corresponding to observational data in accelerating Universe and computing $%
^{ef}\Xi $ for $\ ^{ef}\widehat{L}$ using formulas $\ $%
\begin{equation*}
\ ^{ef}\widehat{\mathbf{T}}_{\alpha \beta }:=\ ^{ef}\widehat{L}\widehat{%
\mathbf{g}}^{\alpha \beta }+2\frac{\delta (\ \ ^{ef}\widehat{L})}{\delta
\widehat{\mathbf{g}}_{\alpha \beta }}\mbox{ and }\ ^{ef}\widehat{\Upsilon }%
_{\ \beta \gamma }:=\varkappa (\ ^{ef}\widehat{\mathbf{T}}_{\alpha \beta }-%
\frac{1}{2}\widehat{\mathbf{g}}_{\alpha \beta }\ ^{ef}\widehat{T})
\end{equation*}
and constraints of type (\ref{2sourc}),
\begin{equation*}
\ ^{ef}\widehat{\Upsilon }_{\alpha ^{\prime }\beta ^{\prime }}=diag[\widehat{%
\Upsilon }_{i};\widehat{\Upsilon }_{a}],\mbox{ for }\widehat{\Upsilon }_{1}=%
\widehat{\Upsilon }_{2}=\ ^{ef}\widetilde{\Upsilon }(x^{k}),\widehat{%
\Upsilon }_{3}=\widehat{\Upsilon }_{4}=\ ^{ef}\widehat{\Upsilon }(x^{k},t).
\end{equation*}%
Then, we can compute $\ ^{ef}\Xi :=\int dt\ ^{ef}\widehat{\Upsilon }(%
\widetilde{\Psi }^{2})^{\ast }.$ Such a problem can be also solved in
inverse form for a given $\widetilde{a}(x^{k},t),$ when $\widetilde{\Psi }$
has to be defined from an integro--differential equation (\ref{aux1}), $%
\widetilde{a}^{2}=\frac{\widetilde{\Psi }^{2}|\ \int dt\ ^{ef}\widehat{%
\Upsilon }(t)(\widetilde{\Psi }^{2})^{\ast }|}{4(\ ^{ef}\Lambda )(\widetilde{%
\Psi }^{\ast })^{2}}$. For cosmological solutions, we can consider $%
\widetilde{a}(x^{k},t)\simeq \widetilde{a}(t)$ and $\widetilde{\Psi }%
(x^{k},t)\simeq \widetilde{\Psi }(t),$ when the generation function $%
\widetilde{\Psi }(t)$ is prescribed to depend only on time--like coordinate $%
t.$ The observable effective scaling factor $\widetilde{a}(t)$ is expressed
as a functional on constant $\ ^{ef}\Lambda ,$ on TMT source $\ ^{ef}%
\widehat{\Upsilon }(t)$ and generating function $\widetilde{\Psi }(t).$ For
instance, variations of $\ ^{ef}\widehat{\Upsilon }(t)$ are determined by
the variation of the second auxiliary 3--index antisymmetric d--tensor field
$\mathbf{B}_{\alpha \beta \gamma }\mathcal{\ }$\ in $^{2}\Phi (\mathbf{B})$
in the formula (\ref{ksiconst}). We adapt and write a similar formula with
"tilde" values in order to emphasize that the values are computed for \ a
prescribed value $\widetilde{a}(t),$%
\begin{equation}
^{2}\Phi (\widetilde{\mathbf{B}})/\sqrt{|\widehat{\mathbf{g}}|}=\ ^{2}%
\widetilde{\chi }=const.  \label{ksiconst1}
\end{equation}%
There are two options to fix a constant $\ ^{2}\widetilde{\chi }:$ the first
one is to chose a function $\widetilde{\Psi }$ and/or to modify $\widetilde{%
\mathbf{B}}$ in the second measure. In general, this is a nonlinear effect
of re--definition of generation functions (\ref{nonltr}) which holds for
generic off--diagonal configurations. We can prescribe finally some small
off--diagonal corrections but the diagonal values will be re--scalled (we
shall keep "tilde " in order to distinguish such values from similar ones
computed from the very beginning using diagonalized equations).

The main conclusion of this subsection is that working with generic
off--diagonal solutions for effective Einstein equations (\ref{densttmt}),
see equivalent formulas (\ref{2mesfeq}), we can chose such generating
functions and effective source that we reproduce in generalized forms the
properties of TMT gravity theories determined by action (\ref{action2}) and
scalar Lagrangians (\ref{scalarlagr}). In the next subsection, we prove that
such models may be generated to have limits to diagonal two flat regions
reproducing accelerating cosmology scenarios.

\subsection{Limits to diagonal two flat regions}

Let us consider in $\ ^{ef}\widehat{L}$ (\ref{eff2}) the approximation
\begin{equation}
\ ^{1}\mathbf{f}(\ ^{1}L+\ ^{1}M,\mu =0)\mathbf{=}\frac{d\mathbf{f}(\widehat{%
\mathbf{R}},\mu =0)}{d\widehat{\mathbf{R}}}\mid _{\widehat{\mathbf{R}}=\
^{1}L+\ ^{1}M}\rightarrow \ \ ^{1}U-\ ^{1}M  \label{starob2}
\end{equation}%
with $\widetilde{\Psi }(t)$ and $\widetilde{a}(t)$ resulting in diagonal
cosmological solutions with effective FLRW metrics. We approximate the
effective potential $\ ^{ef}U$ (\ref{effpot}) for a prescribed constant $\
^{2}\widetilde{\chi }$ by a relation (\ref{ksiconst1}),
\begin{eqnarray*}
\ ^{ef}U&=&\frac{(\ ^{1}U-\ ^{1}M)^{2}}{4\ ^{2}\widetilde{\chi }[\ ^{2}U+\
^{2}M+(\ ^{1}U-\ ^{1}M)^{2}]}\simeq \\
\ ^{ef}\widetilde{U}&=&\left\{
\begin{array}{cc}
^{\lbrack -]}\widetilde{U}=\frac{(\ ^{1}a)^{2}}{4\ ^{2}\widetilde{\chi }[\
^{2}a+\epsilon (\ ^{1}a)^{2}]} & \mbox{ for }\varphi \rightarrow -\infty \\
^{\lbrack +]}\widetilde{U}=\frac{(\ ^{1}M)^{2}}{4\ ^{2}\widetilde{\chi }[\ \
^{2}M+\epsilon (\ ^{1}M)^{2}]} & \mbox{ for }\varphi \rightarrow +\infty%
\end{array}%
\right\vert .
\end{eqnarray*}%
For such diagonal approximations, the $A$-- and $B$--functions can be
computed in explicit form%
\begin{equation*}
\widetilde{\ A}\simeq \left\{
\begin{array}{cc}
^{\lbrack -]}\widetilde{A}=\frac{\ ^{2}a+\frac{1}{2}\ ^{2}b\ ^{1}a}{\
^{2}a+\epsilon (\ ^{1}a)^{2}} &  \\
^{\lbrack +]}\widetilde{A}=\frac{\ ^{2}M}{\ \ ^{2}M+\epsilon (\ ^{1}M)^{2}}
&
\end{array}%
\right\vert \mbox{ and }\widetilde{\ B}\simeq \left\{
\begin{array}{cc}
^{\lbrack -]}\widetilde{B}=-\ ^{2}\widetilde{\chi }\frac{\ ^{2}b/4-\epsilon
(\ ^{2}a+\ ^{2}b\ ^{1}a)}{\ ^{2}a+\epsilon (\ ^{1}a)^{2}} &  \\
^{\lbrack +]}\widetilde{B}=\epsilon \ ^{2}\widetilde{\chi }\frac{\ ^{2}M}{\
\ ^{2}M+\epsilon (\ ^{1}M)^{2}} &
\end{array}%
\right\vert .
\end{equation*}%
Such values reproduce the results of section 3 in \cite{guend2} with two
flat regions of the effective potential $\ ^{ef}\widetilde{U}$ but in our
approach the effective diagonalized metric is of type (\ref{lcconf}) with $%
\check{a}\simeq \widetilde{a}(t)$ for $\eta _{\alpha }\simeq 1.$ This class
of diagonalized solutions but determined by generating functions contain in
explicit form solutions with effective scalar field evolving on the first
flat region for large negative $\varphi $ and describing non--singular
"emergent universes" \cite%
{emergent1,emergent2,emergent3,emergent4,emergent5,emergent6}.

\section{Reproducing Modified Massive Gravity as TMTs and Effective GR}

\label{smassgrfromtmt}The goal of this section is to study solution of
effective Einstein equations (\ref{densttmt}) when the source (\ref{2sourc})
is taken for $~^{m}\widetilde{\Upsilon }=~^{f}\widetilde{\Upsilon }=0$ and $%
~^{m}\Upsilon =~^{f}\Upsilon =0,$ i.e.
\begin{eqnarray*}
\widehat{\Upsilon }_{\alpha ^{\prime }\beta ^{\prime }} &=&diag[\Upsilon
_{i};\Upsilon _{a}],\mbox{ for } \\
\Upsilon _{1} &=&\Upsilon _{2}=\ ^{e\mu }\widetilde{\Upsilon }(x^{k})=\ ^{ef}%
\widetilde{\Upsilon }(x^{k})+~^{\mu }\widetilde{\Upsilon }(x^{k}), \\
\Upsilon _{3}&=&\Upsilon _{4}=\ ^{e\mu }\Upsilon (x^{k},t)=\ ^{ef}\Upsilon
(x^{k},t)+~^{\mu }\Upsilon (x^{k},t),
\end{eqnarray*}%
with a left label $"e\mu "$ emphasizing that such sources are considered for
TMT configurations with a nontrivial mass term $\mu $ but zero matter field
configurations and for a possible quadratic $\epsilon R^{2}$ cosmological
term. We shall chose such N--adapted frames of reference and generating
functions when the TMT gravity model will describe modifications my $\mu
^{2} $ terms for nonholonomic ghost--free configurations and corrections to
scalar curvature (\ref{auxsc}) of type $\mathbf{\check{R}}\simeq \ \widehat{%
\mathbf{R}}+\widetilde{\mu }^{2},$ where
\begin{equation*}
\widetilde{\mu }^{2}\simeq 2~\mu ^{2}(3-tr\sqrt{\mathbf{g}^{-1}\mathbf{q}}%
-\det \sqrt{\mathbf{g}^{-1}\mathbf{q}})
\end{equation*}%
is determined by the graviton's mass $\mu $ and $\mathbf{q}=\{\mathbf{q}%
_{\alpha \beta }\}$ is the so--called non--dynamical reference metric. For
simplicity, we make the assumption that such values can be re--defined to be
constant for certain choices of the generating functions, effective sources $%
\ ^{ef}\Upsilon (x^{k},t),~^{\mu }\Upsilon (x^{k},t)$ and, respective,
nontrivial constants $\ ^{e\mu }\Lambda =\ ^{ef}\Lambda +~^{\mu }\Lambda .$

\subsection{Massive gravity modifications of flat regions}

We can integrate in generic off--diagonal form such TMT systems as
subclasses of solutions (\ref{solnonht}) when
\begin{eqnarray}
ds^{2} &=&\overline{a}^{2}(x^{k},t)[\eta _{1}(x^{k},t)(dx^{1})^{2}+\eta
_{2}(x^{k},t)(dx^{2})^{2}]+  \label{soltmt} \\
&&\overline{a}^{2}(x^{k},t)\widehat{h}_{3}(x^{k},t)[dy^{3}+(\ _{1}n_{i}+\
_{2}n_{i}\int dt\frac{(\widetilde{\Psi }^{\ast })^{2}}{\widetilde{\Psi }%
^{3}(\ ^{ef}\Xi +\ ^{\mu }\Xi )})dx^{i}]^{2}-  \notag \\
&& [dt+\frac{\partial _{i}(\ ^{ef}\Xi +\ ^{\mu }\Xi )}{(\ ^{ef}\Xi +\ ^{\mu
}\Xi )^{\ast }}dx^{i}]^{2},  \notag
\end{eqnarray}%
for
\begin{equation*}
\ ^{ef}\Xi :=\int dt\ ^{ef}\Upsilon (\widetilde{\Psi }^{2})^{\ast },\ ^{\mu
}\Xi :=\int dt\ \ ^{\mu }\Upsilon (\widetilde{\Psi }^{2})^{\ast }.
\end{equation*}
We write $\widetilde{\Psi }\rightarrow \overline{\Psi },$ when the
generating function is chosen to satisfy the conditions
\begin{equation*}
\overline{a}^{2}\widehat{h}_{3}=\omega ^{2}h_{3}=\frac{h_{3}}{|h_{4}|}=\frac{%
\overline{\Psi }^{2}|\ ^{ef}\Xi +\ ^{\mu }\Xi |}{4(\ ^{e\mu }\Lambda )(%
\overline{\Psi }^{\ast })^{2}}.
\end{equation*}
In general, such nonhomogeneous locally anisotropic configurations are with
nontrivial nonholonomically induced canonical d--torsion which can be
constrained to be zero for corresponding subclasses of generating functions
and sources.

We study off--cosmological solutions depending only on time like coordinate
when $\widetilde{a}(x^{k},t)\simeq \widetilde{a}(t)$ and $\widetilde{\Psi }%
(x^{k},t)\simeq \widetilde{\Psi }(t)$ and the generation function $%
\widetilde{\Psi }(t).$ The formula relating variations of $\ ^{e\mu}\Upsilon
(t)$ to the variation of the second auxiliary 3--index antisymmetric
d--tensor field $\mathbf{B}_{\alpha \beta \gamma }$ \ in $\ ^{2}\Phi (%
\mathbf{B}),$ a particular case of (\ref{ksiconst}) is given by
\begin{equation*}
\ ^{2}\Phi (\overline{\mathbf{B}})/\sqrt{|\widehat{\mathbf{g}}|}=\ ^{2}%
\overline{\chi }=\ ^{2}\widetilde{\chi }+\ ^{\mu }\chi =const,
\end{equation*}
where the constant $\ ^{\mu }\chi $ is zero for $\mu =0$ and $|\ ^{\mu }\chi
|\ll |\ ^{2}\widetilde{\chi }|.$ Another assumption is that we can formulate
a TMT\ theory corresponding to "pure" $\mu $--deformations of GR even $%
\epsilon =0.$ The formula (\ref{starob2}) has to be generalized for
nontrivial $\mu ,$ when
\begin{equation*}
\ ^{1}\mathbf{f}(\ ^{1}L+\ ^{1}M+\ ^{\mu }M,\mu )=\frac{d\mathbf{f}(\widehat{%
\mathbf{R}},\mu )}{d\widehat{\mathbf{R}}}\mid _{\widehat{\mathbf{R}}=\
^{1}L+\ ^{1}M}\rightarrow \ \ ^{1}U-\ ^{1}M-\ ^{\mu }M
\end{equation*}
is a version of generalized Starobinsky relation (\ref{starobrelation}),
formulas (\ref{constcond}) and (\ref{auxsc}) and approximation of type $%
\mathbf{\tilde{R}}\simeq \ \widehat{\mathbf{R}}+\widetilde{\mu }^{2}.$

The resulting formulas for effective potential (\ref{effpot}) contain
additional $\mu $--terms
\begin{eqnarray*}
\ ^{e\mu }U &=&\frac{(\ ^{1}U-\ ^{1}M-\ ^{\mu }M)^{2}}{4\ ^{2}\widetilde{%
\chi }[\ ^{2}U+\ ^{2}M+(\ ^{1}U-\ ^{1}M-\ ^{\mu }M)^{2}]} \\
&\simeq &\ ^{e\mu }\overline{U}=\left\{
\begin{array}{cc}
^{\lbrack -]}\overline{U}=\frac{(\ ^{1}a)^{2}}{4(\ ^{2}\widetilde{\chi }+\
^{\mu }\chi )\ [\ ^{2}a+\epsilon (\ ^{1}a)^{2}]} & \mbox{ for }\varphi
\rightarrow -\infty \\
^{\lbrack +]}\overline{U}=\frac{(\ ^{1}M+\ ^{\mu }M)^{2}}{4(\ ^{2}\widetilde{%
\chi }+\ ^{\mu }\chi )[\ \ ^{2}M+\epsilon (\ ^{1}M)^{2}]} & \mbox{ for }%
\varphi \rightarrow +\infty%
\end{array}%
\right\vert .
\end{eqnarray*}%
The $A$-- and $B$--functions can also contain contributions of $\mu $--terms,%
\begin{eqnarray*}
\ \overline{A}&\simeq& \left\{
\begin{array}{cc}
\ ^{[-]}\overline{A}=\frac{\ ^{2}a+\frac{1}{2}\ ^{2}b\ ^{1}a}{\
^{2}a+\epsilon (\ ^{1}a)^{2}} &  \\
\ ^{[+]}\widetilde{A}=\frac{\ ^{2}M}{\ \ ^{2}M+\epsilon (^{1}M+\ ^{\mu
}M)^{2}} &
\end{array}%
\right\vert \mbox{ and } \\
\ \overline{B} &\simeq& \left\{
\begin{array}{cc}
\ ^{[-]}\overline{B}=-\ (\ ^{2}\widetilde{\chi }+\ ^{\mu }\chi )\frac{\
^{2}b/4-\epsilon (\ ^{2}a+\ ^{2}b\ ^{1}a)}{\ ^{2}a+\epsilon (\ ^{1}a)^{2}} &
\\
\ ^{[+]}\overline{B}=\epsilon \ (\ ^{2}\widetilde{\chi }+\ ^{\mu }\chi )%
\frac{\ ^{2}M}{\ \ ^{2}M+\epsilon (^{1}M+\ ^{\mu }M)^{2}} &
\end{array}%
\right\vert ,
\end{eqnarray*}%
when $\ ^{[-]}\overline{A}$ is not modified. We conclude that solutions with
nontrivial generating functions for nontrivial massive gravity terms
modelled as effective TMT\ theories may also describe non--singular
"emergent universes" \cite%
{emergent1,emergent2,emergent3,emergent4,emergent5,emergent6} with
corresponding modifications.

\subsection{Reconstructing off--diagonal TMT and massive gravity
cosmological models}

For the class of solutions (\ref{soltmt}), we show how we can perform a
reconstruction procedure. We introduce a new time coordinate $\widehat{t}$
for $t=t(x^{i},\widehat{t})$ and $\sqrt{|h_{4}|}\partial t/\partial \widehat{%
t},$ and re--defined the scale factor, $\overline{a}\rightarrow $ $\widehat{a%
}(x^{i},\widehat{t}),$ representing the quadratic elements in the form%
\begin{eqnarray}
ds^{2} &=&\widehat{a}^{2}(x^{i},\widehat{t})[\eta _{i}(x^{k},\widehat{t}%
)(dx^{i})^{2}+\widehat{h}_{3}(x^{k},\widehat{t})(\mathbf{e}^{3})^{2}-(%
\widehat{\mathbf{e}}^{4})^{2}],  \label{scdm} \\
\mbox{ for }\eta _{i} &=&\widehat{a}^{-2}e^{\psi },\widehat{a}^{2}\widehat{h}%
_{3}=h_{3},\mathbf{e}^{3}=dy^{3}+\partial _{k}n~dx^{k},\widehat{\mathbf{e}}%
^{4}=d\widehat{t}+\sqrt{|h_{4}|}(\partial _{i}t+w_{i}).  \notag
\end{eqnarray}%
To model small off--diagonal deformations we use a small parameter $%
\varepsilon ,$ $0\leq \varepsilon <1,$ when
\begin{equation}
\eta _{i}\simeq 1+\varepsilon \chi _{i}(x^{k},\widehat{t}),\partial
_{k}n\simeq \varepsilon \widehat{n}_{i}(x^{k}),\sqrt{|h_{4}|}(\partial
_{i}t+w_{i})\simeq \varepsilon \widehat{w}_{i}(x^{k},\widehat{t})
\label{smd}
\end{equation}%
and there are subclasses of generating functions and sources for which $%
\widehat{a}(x^{i},\widehat{t})\rightarrow $ $\widehat{a}(t),\widehat{h}%
_{3}(x^{i},\widehat{t})\rightarrow \widehat{h}_{3}(\widehat{t})$ etc. , see
details for such a procedure from section 5 of \cite{elizaldev} (see also
references therein). The analogous TMT massive gravity theory is taken with
a source $~^{\mu }\widehat{\mathbf{\Upsilon }}_{\mu \nu }$ (\ref{source})
and parametrization $\mathbf{f}(\mathbf{\check{R}})=\ \widehat{\mathbf{R}}+%
\mathbf{S}(~^{\mu }\mathbf{T),}$ for any N--adapted value
\begin{equation*}
~^{\mu }\mathbf{T:=T+}2~\mu ^{2}(3-tr\sqrt{\mathbf{g}^{-1}\mathbf{q}}-\det
\sqrt{\mathbf{g}^{-1}\mathbf{q}}).
\end{equation*}
Introducing values$~^{1}\mathbf{S:=dS/d}~^{\mu }\mathbf{T}$ and $\widehat{H}%
:=\widehat{a}^{\ast }/\widehat{a}$ for a limit $\widehat{a}(x^{i},\widehat{t}%
)\rightarrow \widehat{a}(t)$ with $N_{i}^{a}=\{n_{i},w_{i}(t)\}$ and
effective polarizations $\eta _{\alpha }(t).$

In order to test cosmological scenarios, we consider a redshift $1+z=%
\widehat{a}^{-1}(t)$ for $~^{\mu }T=~^{\mu }T(z)$ by introducing a new
\textquotedblleft shift\textquotedblright\ derivative. For instance, for a
function $s(t)$) $s^{\ast }=-(1+z)H\partial _{z}.$ We can derive TMT massive
modified off--diagonal deformed FLRW equations using formulas (63) and (64)
in \cite{elizaldev}, when {\small
\begin{eqnarray}
3\widehat{H}^{2}+\frac{1}{2}[\mathbf{f}(z)+\mathbf{S}(z)]-\kappa ^{2}\rho
(z) &=&0,  \notag \\
-3\widehat{H}^{2}+(1+z)\widehat{H}(\partial _{z}\widehat{H})-\frac{1}{2}\{%
\mathbf{f}(z)+\mathbf{S}(z)+3(1+z)\widehat{H}^{2} &=&0,  \label{ceq1}
\end{eqnarray}%
} for $\rho (z)\ \partial _{z}\ \mathbf{f}=0.$ We can fix the condition $%
\partial _{z}\ ^{1}\mathbf{S}(z)=0,$ re--scale the generating function in
order to satisfy the condition $\partial _{z}\ \mathbf{f}=0.$ We have
nonzero densities in certain adapted frames of references. Here we note that
the functional $\mathbf{S}(~^{\mu }\mathbf{T)}$ encodes effects of mass
gravity for the evolution of the energy-density when $\rho =\rho
_{0}a^{-3(1+\varpi )}=\rho _{0}(1+z)a^{3(1+\varpi )}$, when for the dust
matter approximation $\varpi $ and $\rho \sim (1+z)^{3}.$ Any FLRW cosmology
can be realized in a corresponding class of $f$--gravity models, which can
be re--encoded as a TMT theories using actions of type (\ref{action1})--(\ref%
{action3}). Let us introduce $\zeta :=\ln a/a_{0}=-\ln (1+z)$ as the
\textquotedblleft e-folding\textquotedblright\ variable \ to be used instead
of the cosmological time $t$ and consider
\begin{equation*}
\ \widehat{\mathbf{\Upsilon }}(x^{i},\zeta )=~\ ^{f}\Upsilon (x^{i},\zeta
)+\ ^{\mu }\Upsilon (x^{i},\zeta )
\end{equation*}
with dependencies on $(x^{i},\zeta )$ of generating functions $\partial
_{\zeta }=\partial /\partial \zeta $ with $q^{\ast }=\widehat{H}\partial
_{\zeta }q$ for any function $q.$

Repeating all computations leading to Eqs.~(2)-(7) in \cite{odintsplb}, in
our approach for $\mathbf{f}(\mathbf{\check{R}}),$ we construct a FLRW like
cosmological model with nonholonomic field equation corresponding to the
first FLRW equation
\begin{equation*}
\mathbf{f}(\mathbf{\check{R}})=(\widehat{H}^{2}+\widehat{H}\ \partial
_{\zeta }\widehat{H})\partial _{\zeta }[\mathbf{f}(\mathbf{\check{R}})]-36%
\widehat{H}^{2}\left[ 4\widehat{H}+(\partial _{\zeta }\widehat{H})^{2}+%
\widehat{H}\partial _{\zeta \zeta }^{2}\widehat{H}\right] \partial _{\zeta
\zeta }^{2}\mathbf{f}(\mathbf{\check{R}})\mathbf{]+}\kappa ^{2}\rho .
\end{equation*}%
We consider an effective quadratic Hubble rate, $\tilde{\kappa}(\zeta ):=%
\widehat{H}^{2}(\zeta ),$ where $\zeta =\zeta (\mathbf{\check{R}}),$ we
write this equation in the form
\begin{equation}
\mathbf{f}=-18\tilde{\kappa}(\zeta )[\partial _{\zeta \zeta }^{2}\tilde{%
\kappa}(\zeta )+4\partial _{\zeta }\tilde{\kappa}(\zeta )]\frac{d^{2}\mathbf{%
f}}{d\mathbf{\check{R}}^{2}}+6\left[ \tilde{\kappa}(\zeta )+\frac{1}{2}%
\partial _{\zeta }\tilde{\kappa}(\zeta )\right] \frac{d\mathbf{f}}{d\mathbf{%
\check{R}}}+2\rho _{0}a_{0}^{-3(1+\varpi )}a^{-3(1+\varpi )\zeta (\widehat{%
\mathbf{R}})}.  \label{flem}
\end{equation}%
For any off-diagonal cosmological models with quadratic metric elements of
type (\ref{scdm}) for redefined $t\rightarrow \zeta $ when a functional $%
\mathbf{f}(\mathbf{\check{R}})$ is used for computing $\widehat{\mathbf{%
\Upsilon }},$ the generating function and respective d--metric and
N--connection coefficients as solutions of certain effective Einstein spaces
for auxiliary connections and effective cosmological constant $\ ^{e\mu
}\Lambda .$ The value $d\mathbf{f/}d\mathbf{\check{R}}$ and higher
derivatives vanish for any functional dependence $\mathbf{f}(\ ^{e\mu
}\Lambda )$ because $\partial _{\zeta }\ ^{e\mu }\Lambda =0.$ We conclude
that the recovering procedure simplifies substantially even in TMT theories
by using re--scaling of generating function and sources following formulas
of type (\ref{nonltr}).

Now we speculate how we can reproduce \textit{the }$\Lambda $\textit{CDM
era. }Using values $\widehat{a}(\zeta )$ and $\widehat{H}(\zeta )$
determined by an off-diagonal quadratic element (\ref{scdm}) and write
analogs of the FLRW equations for ${\Lambda }$CDM cosmology in the form
\begin{equation*}
3\kappa ^{-2}\widehat{H}^{2}=3\kappa ^{-2}H_{0}^{2}+\rho _{0}\widehat{a}%
^{-3}=3\kappa ^{-2}H_{0}^{2}+\rho _{0}a_{0}^{-3}e^{-3\zeta },
\end{equation*}%
for fixed constant values $H_{0}$ and $\rho _{0}.$ The second term in this
formula describes, in general, an inhomogeneous distribution of cold dark
mater (CDM). This allows to compute the effective quadratic Hubble rate and
the modified scalar curvature, $\mathbf{\check{R}}$, \ in the forms $\tilde{%
\kappa}(\zeta ):=H_{0}^{2}+\kappa ^{2}\rho _{0}a_{0}^{-3}e^{-3\zeta }$ and
\begin{equation*}
\mathbf{\check{R}}=3\partial _{\zeta }\tilde{\kappa}(\zeta )+12\tilde{\kappa}%
(\zeta )=12H_{0}^{2}+\kappa ^{2}\rho _{0}a_{0}^{-3}e^{-3\zeta }.
\end{equation*}
The solutions of (\ref{flem}) can be found following \cite{odintsplb} and
\cite{elizaldev} as Gauss hypergeometric functions. We might denote $\mathbf{%
f}=F(X):=F(\chi _{1},\chi _{2},\chi _{3};X),$ where for some constants $A$
and $B$, \
\begin{equation*}
F(X)=AF(\chi _{1},\chi _{2},\chi _{3};X)+BX^{1-\chi _{3}}F(\chi _{1}-\chi
_{3}+1,\chi _{2}-\chi _{3}+1,2-\chi _{3};X).
\end{equation*}%
This is the proof that MGTs and various TMT models can indeed describe ${%
\Lambda}$ CDM scenarios without the need of an effective cosmological
constant because we have effective sources and this follows from the
re--scaling property (\ref{nonltr}) of generic off--diagonal configurations.
The equation (\ref{flem}) transforms into%
\begin{equation}
X(1-X)\frac{d^{2}\mathbf{f}}{dX^{2}}+[\chi _{3}-(\chi _{1}+\chi _{2}+1)X]%
\frac{d\mathbf{f}}{dX}-\chi _{1}\chi _{2}\mathbf{f}=0,  \label{gauss}
\end{equation}%
for certain constants, for which $\chi _{1}+\chi _{2}=\chi _{1}\chi
_{2}=-1/6 $ and $\chi _{3}=-1/2$ where $3\zeta =-\ln [\kappa ^{-2}\rho
_{0}^{-1}a_{0}^{3}(\mathbf{\check{R}}-12H_{0}^{2})]$ and $X:=-3+\mathbf{%
\check{R}}/3H_{0}^{2}.$

Finally, we note that the reconstruction procedure can be performed in
similar form for any MGTs and TMT ones which can modeled, for well--defined
conditions, by effective nonholonomic Einstein spaces.

\section{Results and Conclusions}

\label{sconcl}

\subsection{Modified gravity and cosmology theories with metric Finsler
connections on (co) tangent Lorentz bundles or for nonholonmic Einstein manifolds}

In the present paper and partner works \cite{vexactsol1,
vexactsol1a,vexactsol2,vexactsol3,bubuianu18,bubuianu19,bubuianu20,vacaru20}%
, we follow an orthodox point of view that inflation and accelerating
cosmological models can be elaborated in the framework of effective Einstein
theories via off--diagonal and diagonal solutions for nonholonomic vacuum
and non--vacuum configurations determined by generating functions and
integration functions and constants. Fixing respective classes of such
functions and constants, we can extract various types of modified
gravity--matter theories defined in terms of non--Riemannian volume--forms
(for instance, in a manifestly globally Weyl-scale invariant form) and with
certain modified Lagrange densities of type $f(\mathbf{\check{R}})$
including contributions from the Einstein--Hilbert term $R$, its square $%
R^{2}$, possible massive gravity $\mu $ parametric terms, nonholonomic
deformations etc. The principal results are as follows:

\begin{enumerate}
\item We defined nonholonomic geometric variables for which various classes
of modified gravity theories, MGTs, (in general, with nontrivial
gravitational mass) can be modelled equivalently as respective two measure
(TMT) \cite{guend1a,guend1b,guend1c,guend2,guend3a,guend3b}, bi--connection
and/or bi--metric theories. For well defined nonholonomic constraint
conditions, the corresponding gravitational and matter field equations are
equivalent to certain classes of generalized Einstein equations with
nonminimal connection to effective matter sources and nontrivial
nonholonomic vacuum configurations.

\item We stated the conditions when nonholonomic TMT models encode
ghost-free massive configurations with (broken) scale invariance and such
interactions can modelled by generic off--diagonal metrics in effective
general relativity (GR) and generalizations with induced torsion. Such a
nonholonomic geometric techniques was elaborated in Finsler geometry in
gravity theories and for a corresponding 2+2 splitting we can consider
Finsler like variables and work with so-called FTMT models.

\item We developed the anholonomic frame deformation method \cite%
{bubuianu18,bubuianu19,bubuianu20,vacaru20}, AFDM, in order to generate
off--diagonal, in general, inhomogeneous and locally anisotropic
cosmological solutions in TMT snd MGTs. It was proved that the effective
Einstein equations for such gravity and cosmological models can be decoupled
in general form which allow to construct various classes of exact solutions
depending on generating functions and integration functions and constants.

\item We analysed a very important re--scaling property of generating
functions with association of effective cosmological constants for different
types of modified gravity and matter field interactions which allow to
define nonholonomic variables for which the associated systems of nonlinear
partial differential equations, PDEs, can be integrated in explicit form
when the coefficients of generic off--diagonal metrics and (generalized)
nonlinear and linear connections depend on all space--time coordinates.

\item There were stated conditions on generating functions and effective
sources when zero torsion (Levi--Civita, LC) configurations can be extracted
in general form with possible nontrivial limits to diagonal configurations
in ${\Lambda }$CDM cosmological scenarios, encoding dark energy and dark
matter effects, possible nontrivial zero mass contributions, effective
cosmological constants induced by off--diagonal interactions but finally
constrained nonholonomically to result in nonlinear diagonal effects.

\item A special attention was devoted to subclasses of generic off--diagonal
cosmological solution with effective scalar potentials and two flat regions
and studied limits to diagonal cosmological TMT scenarios investigated in
\cite{guend3a,guend3b}.

\item We studied possible massive gravity modifications of flat regions and
speculated on reconstructing off--diagonal TMT and massive gravity
cosmological models. Via corresponding frame transforms and re--definition
of generating functions and nonholonomic variables, we proved that the same
geometric techniques is applicable in all such MGTs.
\end{enumerate}

Let us explain why it is important to study in different MGTs exact
solutions for off--diagonal and nonlinear gravitational interactions
depending on 2-4 spacetime coordinates and consider possible implications in
modern cosmology. The gravitational and matter field equations in such
theories consist very sophisticate systems of nonlinear PDEs. It was
possible to construct physically important, for instance, black hole and
cosmological solutions for certain diagonal ansatz depending on one
space/time like variable modelling (generalized) Einstein spacetimes with
two and three Killing symmetries or other type high symmetry and asymptotic
conditions. There were two kinds of motivations for such assumptions: The
technical one was that for diagonalizable ansatz the systems of nonlinear
PDEs transform in "more simple" systems of nonlinear ordinary differential
equations, ODEs, which can be integrated in general form. The physical
interpretation of such solutions determined by integration constants is more
intuitive and natural. Nevertheless, a series of problems arisen in modern
acceleration cosmology with dark energy and dark matter effects. It became
clear that standard diagonal cosmological solutions in GR together with
standard scenarios from particle physics and former elaborated cosmological
models can not be applied in order to explain observational cosmological
data. A number of MGTs and new cosmological theories have been proposed and
developed.

Haven chosen mathematically some special diagonalizable ansatz with
prescribed symmetries, we eliminate from consideration another more general
classes of solutions which seem to be important for explaining nonlinear
parametric and nonholonomic off--diagonal interactions. This can be related
to a new nonlinear physics in gravity and particle filed theory which have
not been yet investigated. In the past, there were a number of technical
restrictions to construct such solutions and study their applications but at
present there are accessible advanced numerical, analytic and geometric
methods. In this work, we follow a geometric approach developed in \cite%
{bimmasv1,bimmasv2,vacaru18,vexactsol1,vexactsol1a,vexactsol2,vexactsol3,
bubuianu18,bubuianu19,bubuianu20,vacaru20,rajpootvacaru}, which allow to
construct exact solutions in different classes of gravity and cosmology
theories. Even observational data in modern cosmology can be explained by
almost diagonal and homogeneous models, when possible off--diagonal effects
and anisotropies are very small, this does not constrain us to study only
solutions of associated systems ODEs. For nonlinear gravitational and matter
field systems, a well--defined mathematical approach is to generate (if
possible, exact) solutions in the most general form and then to impose
additional constraints for diagonal configurations. In result, a number of
effects of MGTs and accelerating cosmology can be explained as standard but
off--diagonal nonlinear ones in effective GR. Alternative interpretations in
the framework of TMT and other type theories are also possible.

\subsection{Alternative Finsler gravity theories with metric non-compatible connections}

The referee of this work requested "minimal modification" in order to cite and discuss papers \cite{caponio1,caponio2,caponio3,pfeifer1,hohmann1,lammertzahl} where some alternative Finsler gravity and geometry models are considered. This is a good opportunity for authors which allows them to explain in a more detailed form their approach, geometric methods, and new results on constructing new classes of generic off-diagonal cosmological solutions and elaborating  applications in non-standard particle physics and modified gravity. To comment and compare key ideas and constructions in authors' works  with similar ones from the mentioned alternative geometric and cosmological theories we have to cite additionally the papers \cite{hohmann2,vacaruplb,vacaruapn01,vacarukin1,vacarumon2}, and references therein. We note that in introduction (readers should pay attention to footnote 2) and conclusion sections, and Appendix B, to \cite{vacaru18a} there are provided a number of historical remarks and a review of last 80 years research activity main achievements on Finsler-Lagrange-Hamilton geometry and applications in modern physics, gravity, cosmology, mechanics, information theories.  The axiomatic part was published in \cite{bubuianu18a}.  In just mentioned  works, it is included a study of evolution of main research groups on "Finsler geometry and physics"  in different countries, and formed international collaborations. There were reviewed the results and bibliography of for conventional 20 directions and more than 100 sub-directions of research and publications, of present and other authors, related to Finsler geometry and applications. We also cite as a brief critical review the paper \cite{vacaruplb} and  the monograph \cite{vacarumon2}, (for a collection of works on (non) commutative metric-affine generalized Finsler geometries and nonholonomic supergravity and string theories, locally anisotropic kinetic and diffusion processes, Finsler spinors etc.),  and articles \cite{vacaruapn01,vacarukin1}. Here we summarize and discuss such issues:

\begin{enumerate}
\item In the abstract and introduction, see also subsection 2.2, of this article, it is emphasized that we do not elaborate a typical work on Finsler gravity and
cosmology but rather provide a cosmological work on Einstein gravity and MGTs, TMTs ones, with two measures/ two connections and/or bi-metrics, mass terms, etc., when the constructions are modelled on a Lorentz manifold $V$ of signature (+++-) with conventional
nonholonomic 2+2 splitting. For such theories, the spacetime metrics $g_{\alpha
\beta }(x^{i},y^{a})$ (with $i,j,...=1,2$ and $a,b,...=3,4$) are generic off-diagonal and together with the coefficients of other fundamental geometric objects, depend generically on all spacetime conventional fibred coordinates. Lagrange-Finsler like variables are
introduced on $V$ for "toy" models, when $y^{a}$ are treated similarly to
(co) fiber coordinates on a (co) tangent manifold ($T^{\ast }V$) $TV,$ for a prescribed a
fundamental Lagrange, $L(x,y)$ (or Finsler, for certain homogeneity
conditions $F(x,\beta y)=\beta F(x,y),x=\{x^{i}\}$ etc., for a real constant
$\beta >0,$ when $L=F^{2})$. This states on $V$ a canonical Finsler like N-connection and nonholonomic (co) frames structures, which can be also described in coordinate bases, extracting by additional constraints  the LC-connection or distorting to other linear connections determined by the  same metric structures.   In dual form, we can consider momentum like $p_{a}$-dependencies in $g_{\alpha \beta }(x^{i},p_{a}),$ for a conventional Hamiltonian $H(x,p)$, which can be related to a $L$ via corresponding
Legendre transforms. The reason to introduce such Finsler like and other type nonholonomic variables on a manifold $V,$ or on a tangent bundle $TV$, is that in so-called nonholonomic canonical variables (with hats on geometric objects) the modified Einstein equations (\ref{cdeinst}) can be decoupled and integrated in vary general forms. We have to consider some additional nonholonomic constraints (\ref{lccond}) in order to extract LC-configurations. This is the main idea of the AFDM \cite{bubuianu18,bubuianu19,bubuianu20,vacaru20} which was applied in a series of works for constructing locally anisotropic black hole and cosmological solutions defied by generic off-diagonal metrics and (generalized)
connections in Lagrange-Finsler-Hamilton gravity in various limits of (non)
commutative/ supersymmetric string/ brain theories, massive gravity, TMT
models etc. as we consider in partner works \cite{vexactsol1,
vexactsol1a,vexactsol2,vexactsol3,bubuianu18,bubuianu19,bubuianu20,vacaru20}.

\item One of the formal difficulties in modern Finsler geometry and gravity is that some authors (usually mathematicians) use a different terminology comparing to that  elaborated by physicists in GR, MGTs, TMTs etc. For instance, a theory of "standard static Finsler spaces", with a time like Killing field and/or for static solutions of a type of filed equations in Finsler gravity is elaborated in \cite{caponio1,caponio2,caponio3}. Of course, it is possible always to prescribe a class of static and corresponding smooth class of Finsler generating functions, $F(x,y)$, when  semi-spray, N-connections and d-connections, and certain Finsler-Ricci generalized tensors etc. can be computed for static configurations embedded in locally anisotropic backgrounds. Such constructions can be chosen to be with spherical symmetry. But introducing and computing corresponding "standard static" Sasaki type metrics of type (\ref{dmds}), and their off-diagonal coordinate base equivalents, involving N-coefficients (see the total (phase) spacetime metric (\ref{offn})), we can check that such geometric d-objects (and corresponding canonical d-connection, or LC-connection) do not solve the (modified) Einstein equations (\ref{cdeinst}) if the  data are general ones considered in \cite{caponio1,caponio2,caponio3}. If the d-metric coefficients $g_{\alpha \beta }(x^{i},y^{a})$ are generic off-diagonal with nontrivial N-connection coefficients, such metrics can be only quasi-stationary following the standard terminology in mathematical relativity and MGTs (when coefficients do not depend on time like variable, i.e. $\partial _{t}$ is a Killing symmetry d-vector), but there are nontrivial off-diagonal metric terms because of rotation, N-connections etc. Stationary metrics of type (\ref{dmds}) and/or (\ref{offn}) can be prescribed to describe, for instance, black ellipsoids, which are different from the solutions for  Kerr black holes, BHs, because of a more general Finsler local anisotropy. Static configurations with diagonal metrics of Schwarzschild type BHs can be introduced for some trivial N-connection structures (but in Finsler geometry this is a cornerstone geometric object). For Finsler like gravity theories, there are not proofs of BH uniqueness theorems, and it is not clear if such static configurations (for instance, with spherical symmetry) can be stable.  Such proofs are sketched for black ellipsoids, see details references in \cite{vexactsol1,vexactsol1a,vexactsol2,vexactsol3,bubuianu18,
    bubuianu19,bubuianu20,vacaru20}. So, the existing concepts, definitions, and proofs of "standard" static/ stationary/ cosmological  / stable / nonlinear evolution models etc. depend on the type of postulated principles for respective concepts and theories  of Finsler spacetime.

\item In \cite{pfeifer1,hohmann1,hohmann2},  certain attempts to elaborate models of Finsler spacetime geometry and gravity are considered for some types of  N-connections and chosen classes of Finsler metric compatible and noncompatible d-connections. In many cases, it is considered the  Berwald-Finsler d-connection, which (in general) is noncompatible but can be subjected to certain metrization procedures. Different geometric constructions with non-fixed signature for Hessians and sophisticate causality conditions via semi-sprays and generalized nonlinear geodesic configurations have been proposed and analyzed. In such approaches, there are a series of fundamental unsolved physical and geometric problems for developing such Finsler theories in a some self-consistent and viable physical forms. We point here only some most important issues (for details,  critics, discussions,  and motivation on Finsler gravity principles we cite \cite{vacaru18a,bubuianu18a,vacaruplb,vacarumon2,fmgtcosm1}):

\begin{itemize}
\item For theories with metric noncompatible connections, for instance, of
Chern or Berwald type, there are not unique and simple possibilities to
define spinors, conservation laws of type $D_{i}T^{jk}$, elaborate on supersymmetric and/or noncommutative/ nonassociative generalizations, to consider generalized type classical and quantum symmetries, considering only   Finsler type d-connections proposed by some prominent geometers like E. Cartan, S. Chern, B. Berwald etc., and physically un-motivated (effective) energy-momentum tensors with local anisotropy.

\item Physical principles and nonlinear causality schemes being elaborated on
a base manifold with undetermined lifts, without geometric and physical motivations,  on total bundles depend on the type of Finsler generating functions, Hessians and nonlinear and linear connections are chosen for elaborating geometric and physical models. A Finsler geometry is not a (pseudo/ semi-) Riemannian geometry where all constructions are determined by the metric and LC-connection structures. For instance, certain constructions with cosmological kinetic/ statistical Finsler spacetime in  \cite{hohmann1,hohmann2} are subjected to very complex type conservation laws and nonlinear kinetic/ diffusion equations. Those authors have not cited and did not applied in their works more early locally anisotropic generalized Finsler kinetic/ diffusion / statistical constructions performed for  metric compatible connections studied in \cite{vacaruapn01,vacarukin1,vacarumon2} (N. Voicu were at S. Vacaru seminars in Brashov in 2012, on Finsler kinetics, diffusion and applications in modern physics and information
theory, see also \cite{vacaru20}, but together with her co-authors do not cite, discuss, or apply such locally anisotropic metric compatible and exactly solvable geometric flow, kinetic, geometric thermodynamic theories).

\item Various variational principles, certain versions of Finsler
modified Einstein equations were proposed and developed in \cite%
{pfeifer1,hohmann1,hohmann2} but such theories have been not elaborated on total
bundle spaces, for certain metric compatible Finsler connections. Usually, there were used metric non-compatible Finsler connections,  when it is not possible to elaborate on
certain general methods for constructing exact and parametric solutions of
such nonlinear systems of PDEs, for instance, describint locally anisortopic interactions of modified Finsler-Einstein-Dirac-Yang-Mills-Higgs systems. In the S. Vacaru and co-authors axiomatic approach to relativistic Finsler-Lagrange-Hamilton theories \cite{vacaru18a,bubuianu18a,vacaruplb,fmgtcosm1}, such generalized systems can be studied, for instance, on (co) tangent Lorentz bundles (and on Lorentz manifolds with conventional nonhlonomic fibred splitting), when the AFDM was applied for generating exact and parametric solutions and certain deformation quantization, gauge like etc. schemes were developed.
\end{itemize}

\item In result, authors of \cite{lammertzahl} concluded their work in such a pessimistic form:  "Finsler geometry is a very natural generalisation of pseudo-Riemannian geometry and there are good physical motivations for considering Finsler spacetime theories. We have mentioned the Ehlers-Pirani-Schild axiomatic and also the fact that a Finsler modification of GR might serve as an effective theory of gravity that captures some aspects of a (yet unknown) theory of Quantum Gravity. We have addressed the somewhat embarrassing fact that there is not yet a general consensus on fundamental Finsler equations, in particular on Finslerian generalisations of the Dirac equation and of the Einstein equation, and not even on the question of which precise mathematical definition of a Finsler spacetime is most appropriate in view of physics. We have seen that the observational bounds on Finsler deviations at the laboratory scale are quite tight. By contrast, at the moment we do not have so strong limits on Finsler deviations at astronomical or cosmological scales." In that work, there were not discussed and analyzed the approach developed for Lorentz-Finsler-Lagrange-Hamilton and nonholonomic manifolds developed by authors of this paper  beginning 1994 and published in more than 150 papers in prestigious mathematical and physical journals and summarized in 3 monographs (for reviews, see \cite{vacaru18a,bubuianu18a,vacarumon2}).
\end{enumerate}

S. Vacaru's research group was more optimistic  (than the authors cited above point 4) on obtained results and perspectives of Finsler geometry in physics. Having obtained by 10 NATO, CERN, DAAD research grants, the group  elaborated   an axiomatic approach to Finsler-Lagrange-Hamilton gravity theories using constructions on nonholonomic (co) tangent Lorentz bundles and Lorentz manifolds, with N-connection structure and Finsler like metric compatible connections. They began their activity  almost 40 years ago, see historical remarks, summaries of results and discussions in \cite{vacaru18a,bubuianu18a,vacaruplb,vacarumon2,fmgtcosm1}, with recent developments in \cite{vexactsol1,vexactsol1a,vexactsol2,vexactsol3,bubuianu18,
bubuianu19,bubuianu20,vacaru20}. P. Stavrinos (with more than 40 years research experience on Finsler geometry and applications) and his co-authors also published a series of works on modified Finsler gravity and cosmology theorise involving  tangent Lorentz bundles \cite{fmgtcosm2,stavr3,vacarumon2}.  For such classes of modified Finsler geometric flow and gravity theories, there exists a general  geometric method for  constructing exact and parametric solutions, the AFDM, with self-consistent extensions to noncommutative and nonassociative, supestring and supergravity, Clifford-Finsler etc. theories. Together with papers on deformation and other type quantum Finsler-Einstein-gauge gravity theories, which were elaborated and developed  in more than 20 directions of research on Finsler geometry and applications, this article belongs to an axiomatized and self-consistent direction of mathematical and acceleration cosmology, dark matter and dark energy physics,  involving Finsler geometry methods.

\vskip8pt \textbf{Acknowledgments:} S. V. research develops former programs
partially supported by CERN and DAAD senior fellowships and extended to
collaborations with the California State University at Fresno, the USA, and
Yu. Fedkovych Chernivtsi National University, Ukraine. The main ideas and
results of this work and a respective series of partner papers were
communicated at MG13 Meeting on General Relativity (Stockholm University,
Sweden, 2012), the Particle and Fields Seminar at Physics Department of
Ben-Gurion University of the Negev (Beer Sheva, Israel, 2015) and the IUCSS
Workshop "Finsler Geometry and Lorentz Violation" (Indiana University,
Bloomington, USA, 2017). Authors are grateful to professors E. Guendelman,
A. Kostelecky, N. Mavromatos, Yu. O. Seti, D. Singleton, and M. V. Tkach for
respective support and collaboration. Finally, authors thank an anonymous
referee for his request for "minimal modifications" with discussions of
certain references on alternative Finsler gravity and
cosmology theories.

\end{document}